\documentclass[11pt]{article}
\usepackage{jheppub}
\usepackage{amsmath,amssymb,amsfonts,graphicx,slashed,color, mathtools, amsthm}
\usepackage{comment}
\usepackage{enumerate}
\usepackage{float}
\usepackage{caption}
\usepackage{subcaption}

\newcommand{\be}{\begin{equation}}
\newcommand{\ee}{\end{equation}}
\newcommand{\bea}{\begin{eqnarray}}
\newcommand{\eea}{\end{eqnarray}}
\newcommand{\beas}{\begin{eqnarray*}}
\newcommand{\eeas}{\end{eqnarray*}}
\newcommand{\ba}{\begin{array}}
\newcommand{\ea}{\end{array}}

\newcommand{\nd}{N_{D5}}
\newcommand{\nn}{N_{NS5}}

\title{Finding $AdS^5 \times S^5$ in 2+1 dimensional SCFT physics}

\author[]{Mark Van Raamsdonk,}
\author[]{Chris Waddell}

\affiliation[]{Department of Physics and Astronomy, University of British Columbia,\\
6224 Agricultural Road, Vancouver, B.C.\ V6T 1Z1, Canada.}

\emailAdd{mav@phas.ubc.ca, cwaddell@phas.ubc.ca}

\abstract{We study solutions of type IIB string theory
dual to ${\cal N}=4$ supersymmetric Yang-Mills theory on half of $\mathbb{R}^{3,1}$ coupled to holographic three-dimensional superconformal field theories  (SCFTs) at the edge of this half-space. The dual geometries are asymptotically $AdS^5 \times S^5$ with boundary geometry $\mathbb{R}^{2,1} \times \mathbb{R}^+$, with a geometrical end-of-the-world (ETW) brane cutting off the other half of the asymptotic region of the would-be Poincar\'e $AdS^5 \times S^5$. We show that by choosing the 3D SCFT appropriately, this ETW brane can be pushed arbitrarily far towards the missing asymptotic region, recovering the ``missing'' half of Poincar\'e $AdS^5 \times S^5$. We also show that there are 3D SCFTs whose dual includes a wedge of Poincar\'e $AdS^5 \times S^5$ with an angle arbitrarily close to $\pi$, with geometrical ETW branes on either side.}


\begin{document}

\maketitle

\parskip=10pt

\newpage

\section{Introduction} \label{sec:intro}

\begin{figure}
    \centering
    \includegraphics[width = 80mm]{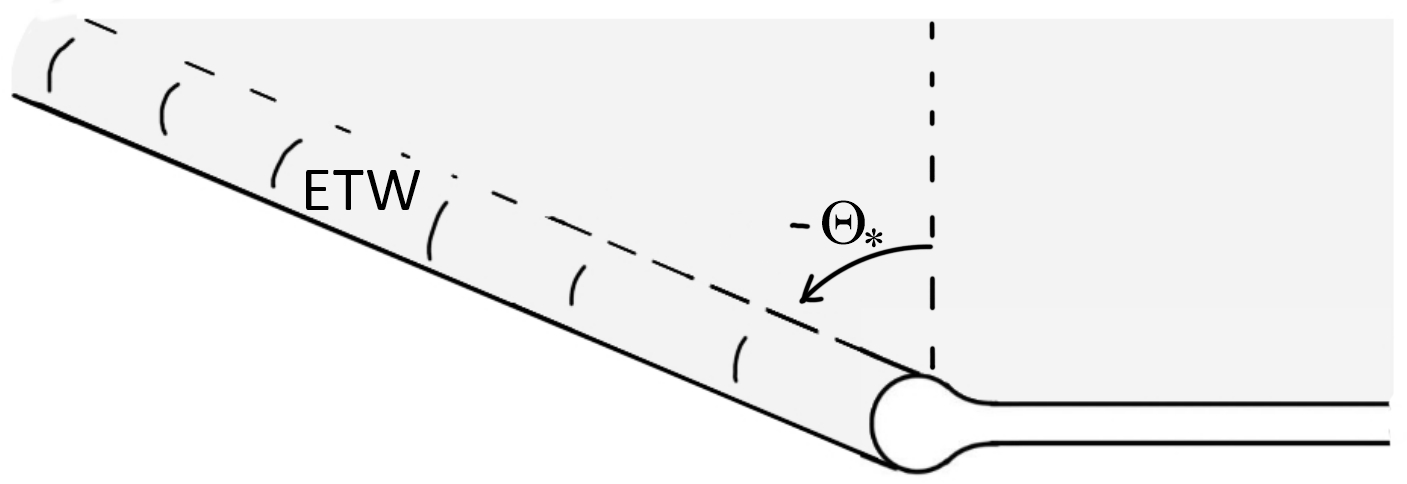}
    \caption{Schematic of geometries dual to ${\cal N}=4$ SYM theory on half of $\mathbb{R}^{3,1}$ coupled to a 3D SCFT at the boundary. The geometry contains a region that approximates a range $\Theta \in (\Theta_*, \pi/2)$ of Poincar\'e $AdS^5 \times S^5$, and an end-of-the-world brane region where the $S^5$ smoothly degenerates. When the 3D boundary SCFT has many more local degrees of freedom than the ${\cal N}=4$ theory, the internal space typically grows to a large volume before pinching off.}
    \label{fig:ETWpic}
\end{figure}

End-of-the-world (ETW) branes arise in many applications of string theory, from model building, to cosmology\footnote{See \cite{Maartens:2010ar} for a review of braneworld cosmology.} \cite{ Cooper:2018cmb, Antonini:2019qkt, VanRaamsdonk:2020tlr, VanRaamsdonk:2021qgv}, to recent studies of black hole evaporation \cite{Almheiri:2019hni,Rozali:2019day, Chen:2020uac, Chen:2020hmv, Geng:2020qvw, Geng:2020fxl,Uhlemann:2021nhu}. 

A particularly interesting case occurs when an ETW brane cuts off the asymptotic region of an asymptotically AdS spacetime \cite{Randall:1999vf}. In this case, gravity can localize on the ETW brane such that over a significant range of scales, gravity on the brane appears to be four dimensional. Such ETW branes can have a microscopic description when the brane intersects the asymptotic boundary of AdS. As explained by Karch and Randall \cite{Karch:2000ct,Karch:2000gx} (see also \cite{Takayanagi:2011zk}), in this case the full system can be dual to a boundary conformal field theory (BCFT). The localization of gravity can arise in the situation where there are many more boundary degrees of freedom than bulk degrees of freedom. 

Often, such ETW branes are considered in bottom-up models where the brane is described as a codimension-one boundary hypersurface with some simple action. In this case, gravity localization can occur when this brane intersects the boundary at a large angle, so that it removes a region $\Theta < \Theta_* = -\pi/2 + \epsilon$ of AdS, where $\Theta$ is the polar angle in Poincar\'e coordinates formed by the radial direction and the field theory direction perpendicular to the CFT boundary. The limit $\epsilon \to 0$ corresponds to the tension of the brane increasing to a critical value. 
 
There are also fully microscopic models which realize ETW brane physics, e.g. \cite{Aharony:2011yc,Assel:2011xz}. In these cases, the ETW brane often corresponds to a region of a higher-dimensional geometry where the internal space degenerates smoothly. In \cite{Bachas:2018zmb}, examples were provided of such microscopic models where gravity is localized to the ETW brane. In this paper, we further study these models, showing that the bulk geometry away from the ETW brane can include a region $\Theta > \Theta_* = -\pi/2 + \epsilon$ of Poincar\'e AdS with arbitrarily small $\epsilon$. That is, we can push the ETW brane arbitrarily far towards the missing asymptotic boundary. 

We further show that there exist solutions with two ETW branes such that the dual contains a region well-approximated by the $-\pi/2 + \epsilon < \Theta <  \pi/2 - \epsilon$ of $AdS$, again with arbitrarily small $\epsilon$.

In the first case, we conclude that the physics of the missing half of the bulk CFT can be reproduced by a set of boundary degrees of freedom. In the second case, the physics of a higher-dimensional CFT can be reproduced by a carefully chosen lower dimensional theory. This is reminiscent of the ``deconstructing dimensions" story \cite{ArkaniHamed:2001ca}.

\subsubsection*{The microscopic setup}

In the microscopic setups we consider, the BCFT is $U(N)$ ${\cal N} = 4$ supersymmetric Yang-Mills theory on $\mathbb{R}^{2,1} \times \mathbb{R}^+$ with boundary physics preserving half supersymmetry and an $OSp(4|4)$ superconformal symmetry.  This boundary physics can generally be understood as a set of boundary degrees of freedom coupled to the ${\cal N} = 4$ fields in some way. These theories arise in string theory from the low energy limit of D3-branes ending on stacks of D5-branes and NS5-branes, with additional D3-branes stretched between the fivebranes. In many cases, the boundary physics can be considered independently and describes a three-dimensional superconformal theory with $OSp(4|4)$ symmetry.

The vacuum states of these field theories on a half-space are dual to known solutions of type IIB supergravity. These solutions have an asymptotically $AdS^5 \times S^5$ asymptotic region whose boundary geometry is half of $\mathbb{R}^{3,1}$. The full geometry has a part that is well approximated by a portion $\Theta > \Theta_*$ of Poincar\'e $AdS^5 \times S^5$, where $\Theta 
\in (-\pi/2,\pi/2)$ is the angle in Poincar\'e coordinates that labels different $AdS^4$ slices and $\Theta = \pi/2$ corresponds to the asymptotic region that is present.\footnote{Here, we assume that $\Theta_*$ is the smallest such angle for which this is true, given some criterion for how closely the geometry should match $AdS^5 \times S^5$.} The remaining part of the geometry can be understood as a geometrical ``end-of-the-world brane'': this is a region of the ten-dimensional geometry where the internal space smoothly degenerates, so that we have a spacetime boundary from the five-dimensional point of view. This ETW brane emerges from the CFT boundary where the SCFT lives. Such geometries are illustrated schematically in Figure \ref{fig:ETWpic}.

For a fixed set of parameters in the ${\cal N}=4$ theory, different choices of the boundary physics (i.e. the choice of 3D SCFT and how this is coupled to the ${\cal N}=4$ theory) give supergravity solutions with the same asymptotically $AdS^5 \times S^5$ region but a different behavior for the ETW brane, and in particular, a different brane angle $\Theta_*$. The main goal of this paper is to show that by choosing the boundary physics appropriately, we can find examples with $\Theta_*$ arbitrarily close to $-\pi/2$. In other words, with the right choice of boundary degrees of freedom, we can, to an arbitrarily good approximation, reproduce the physics of the missing half of the ${\cal N}=4$ theory.

At the level of type IIB supergravity, it is trivial to exhibit families of such solutions that recover all of Poincar\'e $AdS^5 \times S^5$ in a limit. However, the flux quantization conditions of the full type IIB string theory imply that the parameters present in the supergravity solutions cannot be varied continuously, but instead correspond to discrete solutions of a family of nonlinear equations. These parameters correspond to the discrete data used to specify the choice of boundary SCFT to which we couple the ${\cal N}=4$ theory. The non-linear constraints on the supergravity parameters are complicated enough that it is not possible to find a general solution analytically. Nevertheless, we are able to exhibit the existence of sequences of such solutions with the behavior that $\Theta_* \to -\pi/2$. 

On the field theory side, the theories that give $\Theta_* \sim - \pi/2$ correspond to boundary theories with many degrees of freedom. These arise from string theory brane constructions where we have D3-branes ending on stacks of D5-branes and NS5-branes where both $N_{D5}$ and $N_{NS5}$ are taken large. The SCFTs describing these boundary degrees of freedom correspond to the IR limit of quiver gauge theories where the quiver generally has many nodes; we provide some explicit examples below. 

\subsubsection*{Three-dimensional duals to arbitrarily large wedges of $AdS^5 \times S^5$}

For a give choice of boundary physics, we can also consider introducing a second boundary with the same physics (arising from an equivalent configuration of branes) so that supersymmetry is preserved. This theory, now on a strip, will flow to some SCFT in the infrared. The gravity dual for this theory will correspond to a wedge $-|\Theta_*| < \Theta < |\Theta_*|$ of $AdS^5 \times S^5$ with ETW branes on either side. Such solutions were considered in \cite{Bachas:2017rch} and provide a microscopic example of the ``wedge holography'' discussed in \cite{Akal:2020wfl}. Our results in this paper show that the wedge can actually be arbitrarily large, i.e. with an angle that is arbitrarily close to $\pi$. Thus, we can have a 2+1 dimensional theory whose dual geometry contains an arbitrarily large wedge of $AdS^5 \times S^5$.

\subsubsection*{End-of-the-world brane geometries}

The ETW branes in these constructions have a ten-dimensional geometry that was compared by Bachas et al. \cite{Bachas:2018zmb} to a bagpipe. Here, the ``bag'' is a small perturbation to the $AdS^{4} \rtimes M_{6}$ geometry dual to the decoupled 3D SCFT, where $M_{6}$ is a compact internal space. When the SCFT is coupled to the higher-dimensional ${\cal N}=4$ SYM theory, the previously compact internal space $M_{6}$ is perturbed to include a narrow semi-infinite ``pipe'' with the geometry of $S^5$ times a non-compact direction \cite{Bachas:2018zmb}. The perturbation is small since the ${\cal N}=4$ theory has many fewer local degrees of freedom than the SCFT.

The curvature scale of the internal space $M_6$ is generally of the same order of magnitude as the scale $L_4$ describing the non-compact $AdS_4$ geometry of the ETW brane, and these are both much larger than the $AdS^5$ scale $L_5$. The lack of scale separation between the $AdS^4$ scale and the curvature radius of the $M_6$ has been noted in the past \cite{Bachas:2011xa, Bachas:2018zmb}; we provide a direct argument for it in Appendix \ref{app:internal}.

\subsubsection*{Outline}

In the remainder of the paper, we review in Section 2 the field theories that we consider and their gravity duals in type IIB supergravity. In Section 3, we derive conditions on the parameters describing the boundary SCFT such that the dual theories include a region that is well-approximated by a region $\Theta > -\pi/2 + \epsilon$ of $AdS^5 \times S^5$ to an accuracy $\delta$. In Sections 4 and 5, we find explicit examples of sequences of theories (with fixed $g_{\textnormal{YM}}$ and $N$ for the ${\cal N}=4$ theory) that satisfy our conditions with parameters $\epsilon$ and $\delta$ both approaching zero. In Section 6, we describe 3D SCFTs whose duals include arbitrarily large wedges of $AdS^5 \times S^5$ ($|\Theta| < \pi/2-\epsilon$ with arbitrarily small $\epsilon$). We end with a brief discussion in Section 7. 

\section{Background} \label{sec:review}

The field theories we consider and the corresponding supergravity solutions were reviewed in detail in our recent paper \cite{VanRaamsdonk:2020djx}. We refer the reader to Sections 2.2 and 3 of that paper for the details, or to the earlier references \cite{Gaiotto:2008sa, Gaiotto:2008ak} for a discussion of theories with half-maximal supersymmetry  ${\cal N}=4$ on a half-space and \cite{DHoker:2007zhm, DHoker:2007hhe, Aharony:2011yc, Assel:2011xz} for a discussion of the supergravity solutions. Here, we summarize only the basic information that we will use.

The set of supergravity solutions that we discuss take the form of $AdS^4 \times S^2 \times S^2$ fibred over a two-dimensional space $\Sigma$ that we can take to be the positive quadrant of a plane. Explicitly, the metric takes the form
\begin{equation}
    ds^{2} = f_{4}^{2} ds_{\textnormal{AdS}_{4}}^{2} + f_{1}^{2} ds_{S_{1}^{2}}^{2} + f_{2}^{2} ds_{S_{2}^{2}}^{2} +  4 \rho^{2} (dr^2 + r^2 d \theta^2)\: ,
\end{equation}
where $\theta \in [0,\pi/2]$ and $ds_{\textnormal{AdS}_{4}}^{2}$ and $ds_{S_{i}^{2}}^{2}$ are metrics for $AdS^4$ and two-spheres with unit radius. Here, $f_i$ and $\rho$ are functions of $r$ and $\theta$ which are given explicitly in terms of a pair of harmonic functions $h_1,h_2$ on $\Sigma$. 

The general expressions for the harmonic functions corresponding to vacua of ${\cal N}=4$ SYM on a half space with various choices for the boundary physics are given as
\bea
\label{gensol}
h_1 &=&{\pi \ell_s^2 \over 2} {r \cos \theta \over \sqrt{g}} + {\ell_s^2 \over 4}\sum_A {c_A \over  \sqrt{g}} \ln \left( {(r \cos \theta + l_A)^2 +  r^2 \sin^2 \theta \over (r \cos \theta-l_A)^2 + r^2 \sin^2 \theta} \right) \cr
h_2 &=& {\pi \ell_s^2 \over 2} \sqrt{g} r \sin \theta + {\ell_s^2 \over 4}\sum_B d_B \sqrt{g} \ln \left( {r^2 \cos^2 \theta + (r \sin \theta + k_B)^2 \over r^2 \cos^2 \theta + (r \sin \theta-k_B)^2} \right) \; .
\eea
Here the sets $\{l_A\}$ and $\{k_B\}$ represent the locations of poles on the $x$-axis and $y$-axis respectively. These correspond to throats in the ETW brane region of the geometry that are sources of D5-brane flux and NS5-brane flux respectively. The parameters $c_A$ and $d_B$ control the amount of D5 and NS5-brane flux emerging from these throats.

In string theory, the fivebrane flux is quantized; this gives the constraints that 
\begin{equation}
\label{cdN}
        N_{D5}^{(A)} \equiv \frac{1}{\sqrt{g}} c_{A} \in \mathbb{N}^{+}\: , \qquad  N_{NS5}^{(B)} \equiv \sqrt{g} d_{B} \in \mathbb{N}^{+} \: .
\end{equation} 
The throats also have D3-brane flux, and there are additional constraints related to the quantization of this. These are
\begin{equation}
    \label{KLkl}
    L_A = \sqrt{g} l_A + {2 \over \pi} \sum_{B=1} N_{NS5}^{(B)} \arctan{l_A \over k_B} \in \mathbb{N}^{+}\; , \qquad
    K_B = {k_B \over \sqrt{g}} + {2 \over \pi} \sum_{A=1} N_{D5}^{(A)} \arctan{k_B \over l_A} \in \mathbb{N}^{+} \: .
\end{equation}
Here, the integer parameters $L_A$ and $K_B$ can roughly be thought of as the number of units of D3-brane charge per D5-brane associated with the $l_A$ throat or NS5-brane associated with the $k_B$ throat respectively.

\begin{figure}
    \centering
    \includegraphics[width=80mm]{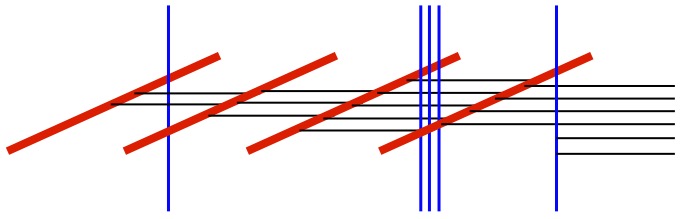}
    \caption{Cartoon of D-brane configuration giving rise to a supersymmetric boundary condition of $\mathcal{N}=4$ SYM; here, D3-branes are black, D5-branes are blue, and NS5-branes are red. This configuration corresponds to linking numbers $\hat{L} = (1, 3, 3, 3, 6)$ and $\hat{K} = (2, 2, 3, 3)$. Removing the semi-infinite D3-branes on the right, we have a brane configuration that gives rise to a 3D SCFT in the infrared.}
    \label{fig:braneconfig2}
\end{figure}

The parameters $(N_{D5}^{(A)},N_{NS5}^{(B)}, L_A, K_B)$ are directly related to the parameters specifying the field theory. The connection is described most easily by referring to the string theory brane constructions from which the field theory arises. It is convenient to define
\beas
(\hat{L}_i) &=& (L_A \; {\rm with \; multiplicity \;} N_{D5}^A) \cr
(\hat{K}_i) &=& (K_B \; {\rm with \; multiplicity \;} N_{D5}^B) \eeas
where both sets are ordered from left to right. 
Then, in the setup of Figure \ref{fig:braneconfig2}, $\hat{K}_i$ is the net number of D3-branes ending from the right on the $i^{\textnormal{th}}$ NS5-brane  \textit{plus} the number of D5-branes to the left of this NS5-brane, and $\hat{L}_i$ is the net number of D3-branes
ending from the right on the $i^{\textnormal{th}}$ D5-brane \textit{plus} the number of NS5-branes to the left of this D5-brane.\footnote{It is sometimes convenient to order the 5-branes such that all NS5-branes occur to the left of all D5-branes; in this case, $\hat{L}_{i}$ is the net number of D3-branes ending on the $i^{\textnormal{th}}$ D5-brane plus the \textit{total} number of NS5-branes, while $\hat{K}_{i}$ is simply the net number of D3-branes ending on the $i^{\textnormal{th}}$ NS5-brane.}

\section{Obtaining a large \texorpdfstring{$AdS^{5} \times S^{5}$}{} region} \label{sec:obtaining}

The solutions dual to $OSp(4|4)$-preserving BCFTs we consider can be thought of as having two general geometrical regions with distinct features:
\begin{itemize}
    \item \textit{Region I:} An asymptotically $AdS^{5} \times S^{5}$ region occurring at large values of the radial coordinate $r \gg l_{A}, k_{B}$ on $\Sigma$, where $O(l_{A}/r)$, $O(k_{b}/r)$ corrections due to the 5-brane throats are small; and
    \item \textit{Region II:} An ``end-of-the-world" brane region at $r \lesssim l_{A}, k_{B}$ where the geometry caps off smoothly except at the locations of the 5-brane throats.  
\end{itemize}

We are interested in considering whether certain allowed choices for the supergravity parameters are able to produce a geometry where region (I) is large and approximates pure $AdS^{5} \times S^{5}$; by ``large", we mean that the $AdS^{5} \times S^{5}$ region extends to Poincar{\'e} angle $\Theta_* \approx -\pi/2$. 

\subsubsection*{Conditions for a large $AdS^5 \times S^5$ region}

Consider the harmonic functions (\ref{gensol}) that determine the metric and other fields. Expanding these in $1/r$, we can write 
\beas
h_1 &=& h_1^{AdS} + {\ell_s^2 \over \sqrt{g}} \left[{1 \over 2} {\sum_A c_A l_A - \sum_B d_B k_B \over r} \cos \theta + \sum_{n=1}^\infty \sum_A c_A \left({l_A \over r}\right)^{2n +1} {\cos((2n+1)\theta) \over 2n+1} \right]\cr
h_2 &=& h_2^{AdS} + \ell_s^2 \sqrt{g} \left[-{1 \over 2} {\sum_A c_A l_A - \sum_B d_B k_B \over r} \sin \theta + \sum_{n=1}^\infty \sum_B d_B \left(-{k_B \over r}\right)^{2n +1} {\sin((2n+1)\theta)\over 2n+1}\right]
\eeas
where 
\bea
\label{h12AdS}
h_1^{AdS} = {L^2 \over 4} {1 \over \sqrt{g}} \cos \theta ({r \over r_0} + {r_0 \over r}) \: , \qquad h_2^{AdS} = {L^2 \over 4} \sqrt{g} \sin \theta ({r \over r_0} + {r_0 \over r})
\eea
are the harmonic functions that give pure $AdS^{5} \times S^{5}$, with AdS length $L$ given by
\be
\label{Nfromkl}
L^4 = 4 \pi \ell_s^4 (\sum_A c_A l_A + \sum_B d_B k_B) = 4 \pi \ell_s^4 N \qquad r_0 = {L^2 \over 2 \pi \ell_s^2} \; .
\ee
For the pure $AdS^{5} \times S^{5}$ solution, the plane $r=r_0$ is an $AdS^4$ slice perpendicular to the boundary that divides the space in half.

We note that for $r \le r_0$, the first term in square brackets will be small compared to the terms in $h^{AdS}$ if and only if
\be
\label{deltacond}
\Delta \equiv \big| \sum_A c_A l_A - \sum_B d_B k_B \big| \ll N \; .
\ee
The ratio $\Delta/N$ gives the fractional size of the corrections (which do not have a significant dependence on $r$ for $r < r_0$). 

The remaining corrections, involving higher powers of $1/r$, become larger (relative to the leading terms) for smaller $r$. It is straightforward to check that these corrections will be small relative to the leading terms provided that $r \gg l_A$ and $r \gg k_B$. For example, when this is true, we have
\be
{\ell_s^2 \over \sqrt{g}} \sum_A c_A {l_A^{2n+1} \over r^{2n+1}} \ll {\ell_s^2 \over \sqrt{g}} \sum_A c_A l_A {1 \over r} \sim {\ell_s^2 \over \sqrt{g}} {N \over r} \approx {L^2 \over \sqrt{g}} {r_0 \over r} \; ,
\ee
where the term on the right is the leading term in $h_1^{AdS}$ in the $r < r_0$ region.

To summarize, we expect that provided the condition (\ref{deltacond}) holds, the solutions will be well approximated by pure $AdS^5 \times S^5$ in a region $r > r_*$ where the coordinate $r$ is much larger than any of the $l_A$ or $k_B$. 
For $h_i = h_i^{AdS}$, the coordinate $r$ is related to the Poincar\'e angle $\Theta$ by \cite{VanRaamsdonk:2020djx}
\be
\frac{r}{r_0} = \tan \left( {\Theta \over 2} + {\pi \over 4} \right) 
\ee
so the geometry includes a region well-approximated by the  $\Theta > \Theta_*$ region of Poincar\'e AdS, where
\be
\Theta_{*} = -{\pi \over 2} + 2 \tan^{-1} {r_* \over r_0} \: .
\ee
In particular, having $\Theta_*$ close to $-\pi/2$ requires $r_* \ll r_0$, which requires \be
\label{klcond}
k_A, l_A \ll \sqrt{N} \; .
\ee
Thus, we have arrived at the two conditions (\ref{deltacond}) and (\ref{klcond}).
In Appendix  \ref{app:cldk_justification},  we provide a more detailed justification that these give solutions with small $\Theta_*$.

\subsubsection*{Satisfying the conditions within string theory}

In the context of type IIB supergravity, it is trivial to find solutions satisfying the constraints (\ref{deltacond}) and (\ref{klcond}) for a given fixed $N$. We are free to take the individual $l_A$ and $k_B$ as small as we like, and then choose $c_A$ and $d_B$ so that 
 \begin{equation}
 \label{Nsum}
        N = \sum_{A} c_{A} l_{A} + \sum_{B} d_{B} k_{B} \: .
    \end{equation} 
and (\ref{deltacond}) is satisfied.

However, in string theory, the solutions obey flux quantization conditions (\ref{cdN}) and (\ref{KLkl}).
Below, we will investigate, for fixed $(g, L_{\textnormal{AdS}})$ (or equivalently fixed parameters $(g_{\textnormal{YM}}, N)$ in the ${\cal N}=4$ theory), the space of parameters $\{l_A,k_B,c_A,d_B\}$ that satisfy both the quantization conditions and the constraints (\ref{deltacond}) and (\ref{klcond}). We will demonstrate discrete families of solutions for which we obtain an arbitrarily large region\footnote{That is, for any $\epsilon > 0$ there exists a solution within the family for which $r_* \ll \epsilon r_{0}$.} of $AdS^{5} \times S^{5}$, approximated arbitrarily well, within the family.

\section{Solutions with single D5-pole and NS5-pole} \label{sec:single}

It is not possible to obtain a large $AdS^{5} \times S^{5}$ region when we have a boundary condition corresponding to a D-brane configuration with only D5-branes or only NS5-branes, since this manifestly violates (\ref{deltacond}) in our constraints. Thus, the simplest possibility is a solution with a single D5-brane throat and a single NS5-brane throat. We consider this case in the present section.

We fix the parameters $N$ and $g$. Then, in terms of the integer parameters $N_{D5}, N_{NS5}$, the relation (\ref{Nfromkl}) and the constraint (\ref{KLkl}) demands that $l, k$ satisfy
\begin{equation} \label{eq:LKSingleStack}
    \begin{split}
        N & = {k \over \sqrt{g}} N_{NS5} + l \sqrt{g} N_{D5}  \\
        L & \equiv \sqrt{g} l + \frac{2}{\pi} N_{NS5} \arctan \left( \frac{l}{k} \right) \in \mathbb{N}^{+} \\
        K & \equiv \frac{k}{\sqrt{g}} + \frac{2}{\pi}  N_{D5} \arctan \left( \frac{k}{l} \right) \in \mathbb{N}^{+} \: ,
\end{split}
\end{equation}

In Appendix \ref{app:lksol}, we show that the allowed $(l, k)$ are in one-to-one correspondence with positive parameters $(N_{D5}, N_{NS5},L,K)$ such that 
\begin{equation}
\label{gcd1}
  G \equiv \textnormal{gcd}(N_{D5}, N_{NS5}) \mid N\: ,
\end{equation}
and
\begin{equation} \label{eq:diophantine1}
    N_{D5} L + N_{NS5} K = N + N_{D5} N_{NS5} \: .
\end{equation}
The latter equation always has at least one solution with positive integers $(L, K)$ provided that (\ref{gcd1}) is satisfied.

In this section, we will understand the space of parameters $(N_{D5}, N_{NS5}, L, K)$ which can realize constraints (\ref{deltacond}) and (\ref{klcond}), and therefore give rise to supergravity solutions with a large region of $AdS^{5} \times S^{5}$. 

The main results of this section are as follows:
\begin{itemize}
    \item If we would like a solution that is well approximated by $AdS^5 \times S^5$ to an accuracy $\delta \ll 1$ in some range $r > \epsilon r_0$ (meaning that $\frac{|cl - dk|}{r_{0}^{2}} \sim \delta^{2}$ ),  it is necessary that $\textnormal{gcd}(N_{D5}, N_{NS5}) \mid N$ and
    \bea
        \nn \gtrapprox \frac{1}{2 \epsilon} \sqrt{gN}   \cr
        \nd  \gtrapprox \frac{1}{2 \epsilon} {\sqrt{N} \over \sqrt{g}} \: .
    \eea
    \item 
    When these are satisfied, the additional condition
    \be
     {\pi \over 8G} \left(\left({g \nd \over \nn} \right) + \left({g \nd \over \nn} \right)^{-1}\right) <  \delta^2 
    \ee
    is sufficient to ensure the existence of suitable $(L, K)$ to give a solution with the desired properties. In particular, if we choose $\nd, \nn$ such that $\textnormal{gcd}(N_{D5}, N_{NS5}) = N$ and $g \nd/\nn = {\cal O}(1)$, the approximation accuracy $\delta$ will be of order $1/\sqrt{N}$. 
    \item We explicitly construct sequences of solutions labeled by a parameter $n \in \mathbb{Z}^+$ with 
    \begin{equation}
        \lim_{n \to \infty} \max\{l(n) , k(n)\} = 0 \: , \qquad \lim_{n \to \infty} |c(n) l(n) - d(n) k(n)| = 0 \: ,
    \end{equation}
    thus obtaining an arbitrarily good approximation to an arbitrarily large $AdS^{5} \times S^{5}$ region for large $n$. 
    For example, in the case of self-dual coupling $g=1$, this occurs for the choice
    \begin{equation}
        N_{D5}(n) = n N \: , \quad N_{NS5}(n) = n N + 2 \: , \quad L(n) = \frac{N}{2}(n - 1) + 1 \: , \quad K(n) = \frac{N}{2} (n + 1) 
    \end{equation}
    (or exchanging $N_{D5} \leftrightarrow N_{NS5}$ and $L \leftrightarrow K$ in these expressions), where $n \in \mathbb{N}^{+}$ is an integer parameter (and we must also require that $n$ is odd if $N$ is odd). 
    More generally, we construct such families for any string coupling $g$ and any choice of relative scaling $z N_{D5} \sim N_{NS5}, z \in \mathbb{R}^{+}$. 
\end{itemize}

\subsection{Necessary conditions for solutions with large \texorpdfstring{$AdS^{5} \times S^{5}$}{} region}

Suppose we would like a solution that is well approximated by $AdS^5 \times S^5$ to an accuracy $\delta$ in some range $r > \epsilon r_0$. Then according to the conditions (\ref{deltacond}) and (\ref{klcond}) we require that
\bea
& l < \epsilon \sqrt{N} \label{Cl}\\
& k < \epsilon \sqrt{N} \label{Ck}\\
& {1 \over \sqrt{N}}\left|{k \nn \over \sqrt{g}} - l \nd \sqrt{g}\right|^{1 \over 2} < \delta \; . \label{Cd}
\eea
Recalling that
\be
{k \over \sqrt{g}} \nn + l \sqrt{g} \nd = N \; , \label{klN}
\ee
we may combine (\ref{Cd}) and (\ref{klN}) to find that
\bea
{N \over 2} ( 1 - \delta^2) &<& {k \over \sqrt{g}} \nn < {N \over 2} ( 1 + \delta^2) \cr
{N \over 2} ( 1 - \delta^2) &<& l \sqrt{g} \nd < {N \over 2} ( 1 + \delta^2) \; .
\label{klbound}
\eea
Combining these with (\ref{Cl}) and (\ref{Ck}), we see that 
\bea
\nn  > \sqrt{gN}\frac{1}{2 \epsilon} \left( 1 - \delta^{2} \right) \cr
\nd  > \sqrt{N \over g} \frac{1}{2 \epsilon} \left( 1 - \delta^{2} \right)   \; .
\label{NDcond}
\eea
Consequently, we see that \textit{both} $\nd$ and $\nn$ must be sufficiently large for (\ref{deltacond}) and (\ref{klcond}) to simultaneously be satisfied, in addition to the previous requirement $G \mid N$. Notably, this implies that if we would like to construct a family of solutions which can achieve an arbitrarily large $AdS^{5} \times S^{5}$ region, then we will need to take both $N_{D5}$ and $N_{NS5}$ to be increasingly large within this family.

\subsection{Sufficient conditions for solutions with large \texorpdfstring{$AdS^{5} \times S^{5}$}{} region}

Given $N_{D5}, N_{NS5}$ satisfying $G \mid N$ and (\ref{NDcond}), we will now investigate the additional conditions which guarantee a choice of $(l, k)$ in the range (\ref{klbound}) for which $L$ and $K$ are integers.   

For $\nd$ and $\nn$ satisfying constraints (\ref{NDcond}) and $G \mid N$, we have from (\ref{klbound}) that
\be
{k \over l} \in g{\nd \over \nn}[1 - 2 \delta^2, 1 + 2 \delta^2] \; .
\label{klres}
\ee
Using (\ref{eq:LKSingleStack}) together with (\ref{klbound}) and (\ref{klres}), we have that
\bea
L &\approx& L_0 = {N \over 2 \nd} + {2 \over \pi} \nn \arctan \left( {\nn \over g \nd} \right) \cr
K &\approx& K_0 = {N \over 2 \nn} + {2\over \pi} \nd \arctan \left({g \nd \over \nn} \right) \;. \label{LKapprox}
\eea
More precisely, taking into account the allowed range of $l$ and $k/l$, $L$ must lie in a range of values with half width
\be
\Delta L = 2 \delta^2 \left[{N \over 4 \nd} + {2 \over \pi} {g \nd \over 1 + \left({g \nd \over \nn} \right)^2} \right] \; .
\label{Lrange}
\ee
We can show that the second term here is larger when (\ref{NDcond}) is satisfied, so we can take the range as
\be
\Delta L \approx 2 \delta^2 \left[{2 \over \pi} {g \nd \over 1 + \left({g \nd \over \nn} \right)^2} \right] \; .
\ee
We need the range $[L_0 -\Delta L, L_0 + \Delta L]$ to be large enough to contain an integer value. More specifically, we need a value for which $K - N_{D5} = (N - L \nd)/\nn$ is also an integer. This requires that $G \mid N$, in which case, a range of $L$ of length $\nn/G$ will lead to 
at least one integer value of $K$.

Thus, for fixed $g$ and $N$, and some chosen $\nd$ and $\nn$ satisfying the constraints (\ref{NDcond}) and that $G \mid N$, we will get a solution provided that the range  (\ref{Lrange}) is at least $\nn/G$; that is, it should be sufficient that
\be
{1 \over G} < \delta^2 \left[{8 \over \pi} {\left({g \nd \over \nn} \right) \over 1 + \left({g \nd \over \nn} \right)^2} \right] \; ,
\ee
or 
\be
{\pi \over 8G} \left(\left({g \nd \over \nn} \right) + \left({g \nd \over \nn} \right)^{-1}\right) <  \delta^2 \; .
\ee
Since the term in brackets is larger than or equal to $2$ and $G < N$, we expect that our sufficient condition can be satisfied provided that $\delta$ is at least $1 / \sqrt{N}$. However, we will see below that for fixed $N$, arbitrarily small values of $\epsilon$ and $\delta$ are possible for carefully chosen parameters.

\subsection{One-parameter families with arbitrarily large \texorpdfstring{$AdS^{5} \times S^{5}$}{} region} \label{sec:oneparam}

For simplicity, we will begin with the case of self-dual coupling $g=1$.
We consider a sequence of parameters labeled by $n \in \mathbb{N}^{+}$ (and further imposing that $n$ is odd for odd $N$ to satisfy (\ref{gcd1})), defining
\begin{equation}
    N_{D5} = n N \: , \quad N_{NS5} = n N + 2 \: , 
\end{equation}
and 
\begin{equation} \label{eq:alphaND5NNS5}
    L = \frac{N}{2} (n - 1) + 1 \: , \quad K = \frac{N}{2} (n + 1) \: ,
\end{equation}
or alternatively, using the same expressions but with $N_{D5} \leftrightarrow N_{NS5}$ and $L \leftrightarrow K$.
In this case, we can check that (\ref{gcd1}) and (\ref{eq:diophantine1}) are satisfied, so our results in Appendix \ref{app:lksol} show that there will be a unique choice $(l, k)$ satisfying (\ref{eq:LKSingleStack}).

For large $n$, we can write this solution perturbatively as
\begin{equation} \label{eq:pertlk}
\begin{split}
    l & = \frac{1}{2 n} - \frac{1}{2 n^{2}} \left( \frac{\pi}{4} + \frac{1}{N} \right) + {\cal O}(n^{-3}) \\ k & = \frac{1}{2 n} + \frac{1}{2 n^{2}} \left( \frac{\pi}{4} - \frac{1}{N} \right)  + {\cal O}(n^{-3}) \; .
\end{split}
\end{equation}
From these, we find that
\begin{equation}
    |c l - d k| = \frac{1}{n} \left( 1 + \frac{\pi N}{4} \right) + {\cal O}(n^{-2}) \: .
\end{equation}
so we can indeed make $\max\{l, k\}$ and $|cl - dk|$ arbitrarily small within this particular class of solutions, by choosing sufficiently large $n$. 
Thus, we can have an arbitrarily large region of $AdS^5 \times S^5$ arbitrarily well-approximated by our solution.

\begin{figure}[h]
\centering
\begin{subfigure}{.5\textwidth}
  \centering
  \includegraphics[height=7cm]{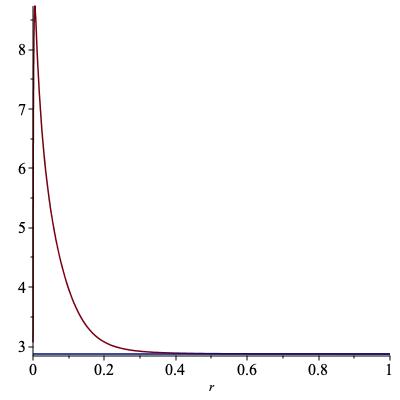}
  \caption{$\ln f_{1}^{2}\left(r, \frac{\pi}{4} \right)$ versus $r$}
  \label{fig:f1_2_single}
\end{subfigure}%
\begin{subfigure}{.5\textwidth}
  \centering
  \includegraphics[height=7cm]{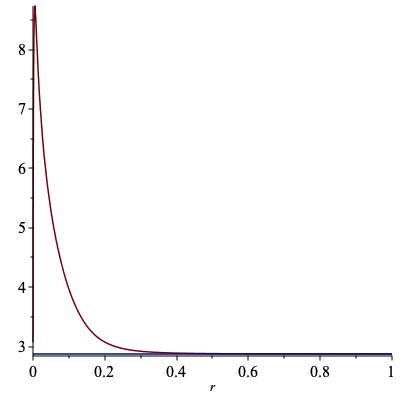}
  \caption{$\ln f_{2}^{2}\left(r, \frac{\pi}{4} \right)$ versus $r$}
  \label{fig:f2_2_single}
\end{subfigure} 
\\
\begin{subfigure}{.49\textwidth}
  \centering
  \includegraphics[height=7cm]{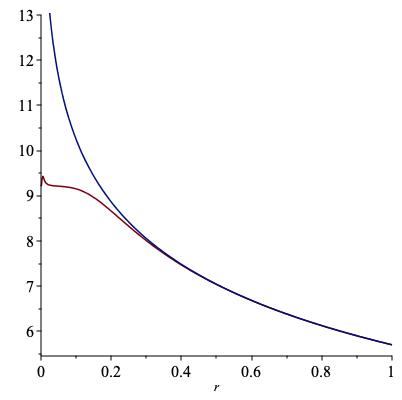}
  \caption{$\ln f_{4}^{2}\left(r, \frac{\pi}{4} \right)$ versus $r$}
  \label{fig:f4_2_single}
\end{subfigure}
\begin{subfigure}{.49\textwidth}
  \centering
  \includegraphics[height=7cm]{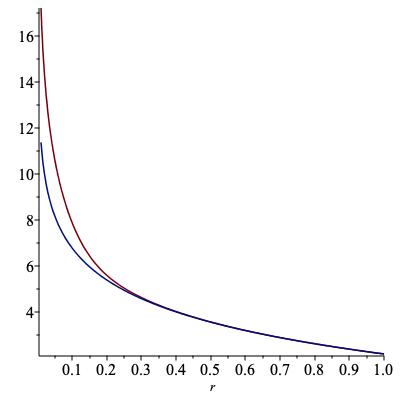}
  \caption{$\ln \rho^{2}\left(r, \frac{\pi}{4} \right)$ versus $r$}
  \label{fig:rho_2_single}
\end{subfigure}
\caption{In these figures, we are taking $g=1, \ell_{\textnormal{s}} = 1, N = 100$. The metric functions shown in red are for the case $(c, d, l, k) = (10^{4}, 10^{4} + 2, 4.96 \times 10^{-3}, 5.04 \times 10^{-3})$ (namely $n=100$ in our family of solutions), while the metric functions shown in blue are for pure $AdS^{5} \times S^{5}$. Note that in this case $r_{0} \approx 5.64$. } 
\label{fig:metricfunctions_single}
\end{figure}
\begin{figure}[h]
\centering
\begin{subfigure}{.5\textwidth}
  \centering
  \includegraphics[height=6cm]{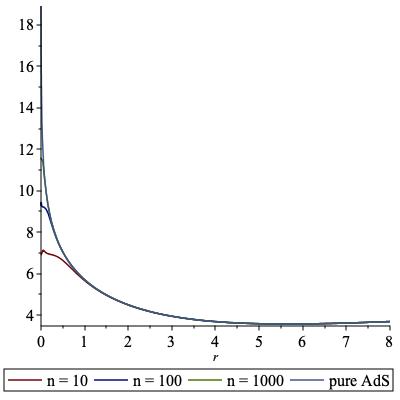}
  \caption{$\ln\left(f_{4}^{2}\left( r, \frac{\pi}{4} \right)\right)$ versus $r$ on $r \in (0, 8)$}
  \label{fig:f42_1}
\end{subfigure}%
\begin{subfigure}{.5\textwidth}
  \centering
  \includegraphics[height=6cm]{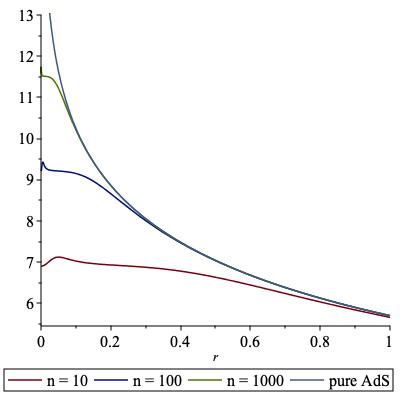}
  \caption{Close-up: $\ln\left(f_{4}^{2}\left( r, \frac{\pi}{4} \right)\right)$ versus $r$ on $r \in (0, 1)$}
  \label{fig:f42_2}
\end{subfigure}
\caption{In these figures, we are taking $g=1, \ell_{\textnormal{s}} = 1, N = 100$. The metric functions shown correspond to the indicated values of $n$ in the family of solutions above, as well as the case of pure $AdS^{5} \times S^{5}$. }
\label{fig:f42comparison}
\end{figure}

To emphasize that these choices of parameters indeed give rise to a large $AdS^{5} \times S^{5}$ region, we show in Figures \ref{fig:metricfunctions_single} and \ref{fig:f42comparison} the metric functions obtained for particular choices of these parameters, as well as the metric functions of $AdS^{5} \times S^{5}$ for reference. 
We find that these metric functions agree to good approximation for $r$ above some $r_*$ which becomes small as the parameter $n$ is taken to be large.

\subsubsection*{General construction of one-parameter families} 
Next, we consider a more general case where the string coupling takes the form
\be
\label{defg}
g = m \cot \left( {\pi \over 2} {a \over b}\right)
\ee
where $m \in \mathbb{Z}^+$ and $a < b$ are relatively prime. The set of such string couplings is dense in $[0,\infty)$. Taking $(\alpha, \delta)$ to be any solution to the Diophantine equation\footnote{A simple explicit case is to take $b = a+1$ (so that $g = m \tan(\pi/(2b))$), $\alpha = N$, and $\delta =0$.}
\be
(b-a) \alpha - b \delta = N
\ee
we define a sequence\footnote{Note that different choices for $(\alpha,\delta)$ lead to the same sequence with a redefinition of $n$.}
\beas \label{eq:generalseq}
\nd(n) &=& bn + \alpha \cr
\nn(n) &=& m (bn + \alpha) - b \cr
L(n) &=& am n - a + m (\alpha - \delta) \cr
K(n) &=& (b-a)n + \delta
\eeas
We can also consider a similar sequence with the replacements $\nd \leftrightarrow \nn, L \leftrightarrow K, g \leftrightarrow 1/g$. This choice is motivated in Appendix \ref{sec:generalfam}.

For these choices, it is straightforward to check that (\ref{eq:diophantine1}) is satisfied. Also, constraint $\sqrt{g} N_{D5} l + \frac{1}{\sqrt{g}} N_{NS5} k = N$ implies that both $l$ and $k$ are at most $O\left( n^{-1} \right)$, so these go to zero in the limit $n \to \infty$. Finally, we need to verify that $|c l - d k|$ also vanishes in this limit.

From the definitions of $L(n)$, $\nn(n)$, and $g$, we see that
\be
{L(n) \over \nn(n)} = \frac{a}{b} + O \left( n^{-1} \right) = \frac{2}{\pi} \arctan(m/g) + O \left( n^{-1} \right)
\ee
The equations (\ref{eq:LKSingleStack}) yield
\begin{equation}
        \frac{L(n)}{N_{NS5}(n)}
         = \frac{2}{\pi} \arctan(l/k) + O \left(n^{-2} \right) \: .
\end{equation}
Thus, we have 
\begin{equation}
    l/k = m/g + O \left( n^{-1} \right) \: .
\end{equation}
It follows that
\begin{equation}
        |cl - dk| = \big| \left( \sqrt{g} b n \right) \left( \frac{km}{g} + O\left( n^{-2} \right) \right)  - \left( \frac{1}{\sqrt{g}} b m n \right) k \big| = O \left( n^{-1} \right) \: ,
\end{equation}
as desired. Thus, an arbitrarily large region of $AdS^5 \times S^5$ becomes arbitrarily well approximated for solutions corresponding to large enough $n$. 

The construction so far applies to a particular dense set of string couplings of the form (\ref{defg}), and leads to a scaling of parameters
\begin{equation}
    N_{D5} \sim m N_{NS5} \; ,
\end{equation}
where $m$ is an integer. In Appendix \ref{sec:generalfam}, we generalize the construction to arbitrary real string coupling and find families of solutions that exhibit a more general scaling $N_{NS5} \sim z N_{D5}$ for arbitrary $z > 0$.

For general $z$, conditions (\ref{deltacond}) and (\ref{klcond}) then fix the scaling for the linking numbers to be
\begin{equation}
    \frac{L}{N_{NS5}} \sim \frac{2}{\pi} \arctan(z / g) \: , \qquad \frac{K}{N_{D5}} \sim \frac{2}{\pi} \arctan(g / z) \: .
\end{equation}

\subsection{Field theory interpretation for solution families approaching $AdS^5 \times S^5$} \label{sec:fieldtheory}

We would now like to understand from the field theory perspective what boundary physics for the ${\cal N}=4$ SYM theory gives rise to the solutions with arbitrarily large regions of $AdS^5 \times S^5$ ($\Theta_*$ arbitrarily close to $-\pi/2$). In each case, we are coupling the ${\cal N}=4$ SYM theory on a half space to a particular 3D SCFT\footnote{ We recall that the general $OSp(4|4)$-invariant boundary condition of this theory can be specified by a triple $(\rho, H, \mathcal{B})$ \cite{Gaiotto:2008sa, Gaiotto:2008ak}; here, $\rho : \mathfrak{su}(2) \rightarrow \mathfrak{g}$ is a homomorphism into the Lie algebra of the gauge group (in our case $U(N)$) which specifies the ``Nahm pole" boundary condition for the scalars in the bulk 4D hypermultiplet, $H$ is the residual symmetry group at the field theory boundary, and $\mathcal{B}$ is the 3D SCFT coupled at the boundary. For the boundary conditions in the one-parameter families that we are currently considering, we are imposing a simple Dirichlet boundary condition on the bulk hypermultiplet (and a Neumann condition on the 4D vector multiplet), and there is no reduction in gauge symmetry; our boundary conditions are then entirely specified by the SCFT $\mathcal{B}$.} that can be understood as arising from the low-energy physics of a particular brane configuration in string theory, or as the IR limit of a quiver gauge theory.

To understand the brane construction corresponding to parameters $(N_{D5}, N_{NS5}, L, K)$, we note that the parameters $\hat{L}_i$ introduced in Section 2 are simply $L$ with multiplicity $\nd$, while the parameters $\hat{K}_i$ are $K$ with multiplicity $\nn$. From the relation between these parameters and the brane configuration, we can check that this set corresponds to having $\nn$ NS5-branes which we can initially think of as being separated along a direction $x^3$ (the direction in which the D3-branes are semi-infinite), with a stack of $\nd$ D5-branes between the $L$th and $(L+1)$st NS5-brane from the left. We additionally have $n_i$ D3-branes streteched between the $i$th and $(i+1)$st NS5, where
\be
n_i = \left\{ \ba{ll} i K & i \le L \cr
iK - \nd (i-L) & i > L \ea \right.
\ee
To the right of the final NS5-brane, we have the $N$ semi-infinite D3-branes.

\begin{figure}[t]
\centering
\includegraphics[height=3cm]{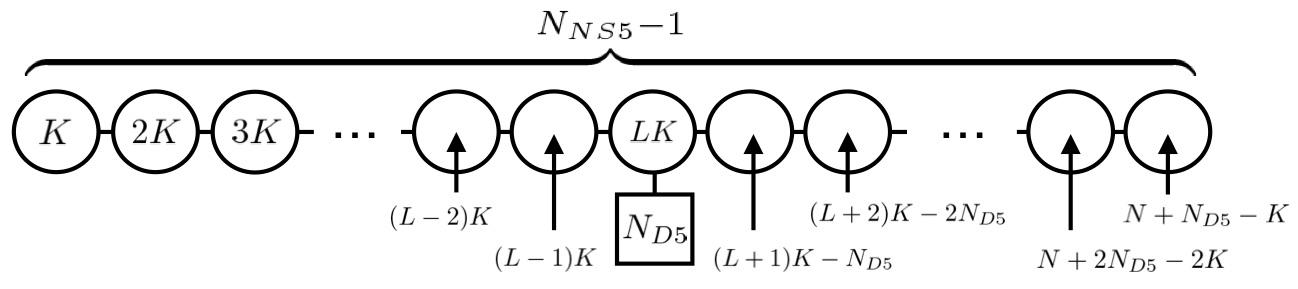}
\caption{General form of quiver gauge theory which corresponds to the field theory boundary conditions determined by $(N_{D5}, N_{NS5}, L, K)$. 
}
\label{fig:LKQuiver}
\end{figure}

Stripping off the semi-infinite D3-branes gives a brane setup whose low-energy physics is a SCFT that corresponds to the IR limit of the quiver gauge theory shown in Figure \ref{fig:LKQuiver}. Such a quiver consists of $N_{NS5}-1$ nodes, with $N_{D5}$ fundamental hypermultiplets coupled to the $L^{\textnormal{th}}$ node. For nodes to the left of the $L^{\textnormal{th}}$ node, the gauge group rank increases in increments of $K$ as we read the quiver from left to right; for nodes to the right, the gauge group rank decreases in increments of $N_{D5} - K$.

So far, this construction is completely general within boundary conditions involving a single D5-brane throat and a single NS5-brane throat; we now restrict to boundary conditions within the families 
considered in this section. For the one-parameter family introduced at the beginning of Section \ref{sec:oneparam} (with $g=1$ and $z=1$), we see that the corresponding quiver is approximately ``left-right symmetric" for large $n$; given that our family has $\frac{L}{N_{NS5}} \approx \frac{1}{2}$ for large $n$, the hypermultiplets are coupled to a single node which is roughly in the middle of the quiver, after which the gauge group rank decreases in increments of $N_{D5} - K$, where $\frac{N_{D5} - K}{K} \approx 1$ for large $n$. 
More generally, we find that, if we parametrize the quiver by its length $N_{NS5} - 1 \approx N_{NS5}$, then we will have $N_{D5} \approx \frac{1}{z} N_{NS5}$ fundamental hypermultiplets coupled to a node whose placement in the quiver grows proportionally to the length of the quiver to enforce the ratio $\frac{L}{N_{NS5}} \approx \frac{2}{\pi} \arctan (z / g)$. 
In particular, we note that in the case of small coupling $g \ll z$, the fundamental hypermultiplets will be roughly at the right end of the quiver, while in the case of large coupling $g \gg z$ they will be at the left end. 

The fact that all the hypermultiplets are attached to the same gauge group factor (or that the D5-branes in the brane construction come in a single stack) is an artifact of our simplifying assumption that the harmonic functions leading to the supergravity solution have only a single D5-brane pole and a single NS5-brane pole. We expect that there are many other choices with additional poles that lead to more general quivers but still give $\Theta_* \to -\pi/2$ in a limit. In Appendix \ref{app:pert}, we will verify that such cases can be obtained by small deformations of the boundary conditions in this section. In particular, we construct examples where we couple in additional hypermultiplets to an additional node of the quiver; this corresponds to adding in additional D5 and NS5-brane poles. We also consider deforming our single-pole boundary conditions by coupling the corresponding quivers to an additional small quiver at the left endpoint.
In both of these contexts, we find more general sequences of solutions that still yield $\Theta_* \to -\pi/2$. 
We will consider a further generalization with multiple D5-brane poles in the following section.

\section{Solutions with multiple poles}

In this section, we consider a more general case where we still have only a single NS5-brane pole in $h_2$ at location $y=k$ with multiplicity $\nn$, but we allow arbitrary numbers of D5-brane poles in $h_1$ at locations $x = l_A$. 

These poles will correspond to some linking numbers $K$ with multiplicity $\nn$ and $\nd$ linking numbers $\{L_A\}$, such that
\be
\label{diop1}
 \nn K + \sum_A L_A = N + \nn \nd \; .
\ee
Given linking numbers satisfying this, the corresponding pole locations $k$ and $l_A$ must satisfy
\bea
L_A &=& \sqrt{g} l_A + {2 \over \pi} \nn \arctan {l_A \over k} \cr
K &=& {k \over \sqrt{g}} + {2 \over \pi} \sum_A \arctan{k \over l_A} \; .
\label{KLeqs}
\eea
We can determine $k$ and $l_A$ as follows. Defining 
\be
F_k(x) = \sqrt{g} x + {2 \over \pi} \nn \arctan{x \over k}
\ee
and noting that for any $k$, $F_k$ is a monotonic map from $[0,\infty)$ to $[0,\infty)$, we have that
\be
l_A = F_k^{-1}(L_A) \: .
\ee
The actual value of $k$ is determined by solving\footnote{We note that each term on the right is monotonically increasing with $k$, and the entire right side increases monotonically from a value less than $N$ for $k=0$ to infinity for $k=\infty$, so there will be a unique solution.}
\be
N = {k \over \sqrt{g}} \nn + \sqrt{g} \sum_A F_k^{-1}(L_A) \; .
\ee

To see which linking numbers satisfy our conditions for having a $\Theta_*$ close to $-\pi/2$, we note that the requirements that
\be
\label{condA}
\sqrt{g} \sum_A l_A + {k \over \sqrt{g}} \nn = N \; .
\ee
(which follows from the first three equations of this section) and our condition
\be
\label{condB}
|\sqrt{g} \sum_A l_A - {k \over \sqrt{g}} \nn | \ll N
\ee
require that both terms in each expression are close to $N/2$ so
\be
\label{lAcond}
\sqrt{g} \sum_A l_A \approx {N \over 2}
\ee
and
\be
\label{k0def}
k \approx k^{(0)} \equiv  {\sqrt{g} N \over 2 \nn} \; .
\ee
In order that $k \ll \sqrt{N}$, the latter condition implies 
\be
\label{nncond}
\nn \gg \sqrt{g N} \; .
\ee 
Then the $l_A$ are approximately related to $L_A$ by
\be
\label{l0def}
l_A \approx l_A^{(0)} \equiv F_{k^{(0)}}^{-1}(L_A) \: .
\ee
The condition $l_A \ll \sqrt{N}$ gives that
\be
\label{L0cond}
F_{k^{(0)}}^{-1}(L_A) \ll \sqrt{N} \; .
\ee
From the condition (\ref{lAcond}), we have 
\be
\label{Lcond}
\sqrt{g} \sum_A F_{k^{(0)}}^{-1}(L_A) \approx {N \over 2} \; .
\ee
Since each $l_A = F_{k^{(0)}}^{-1}(L_A)$ in the sum is required to be much less than $\sqrt{N}$ but also greater than or equal to $F_{k^{(0)}}^{-1}(1) \sim \pi \sqrt{g}N / (4 \nn^2)$, we note that the number of D5-brane poles (including multiplicity) must satisfy
\be
\label{ndcond}
{1 \over 2}  \sqrt{N \over g} \ll \nd < {2 \nn^2 \over \pi g} \; .
\ee
Our choice of the $L_A$ must be such that
\be
K = N_{D5} + {N - \sum_A L_A \over \nn}
\ee
is an integer. To see when this is possible, we note that for $L_A \ll \nn$, $F$ is linear and 
\be
\label{Lapprox}
l_A = F_{k^{(0)}}^{-1}(L_A) \approx {\pi \over 4} {\sqrt{g} N  \over \nn^2} L_A \; .
\ee
Thus, adding an additional pole with $L=1$ or varying one of the $L_{A}$ by 1 leads to a change in the left side of (\ref{Lcond}) of 
\be
{\pi \over 4} {g N  \over \nn^2} \ll 1\; .
\ee
Given any set of $L$s, changing the sum by an amount less than $\nn$ will be enough to give an integer  $K$. If we add or change $L$s in the linear regime of $F$, the change in $\sqrt{g} \sum l_A$ will be less than
\be
{\pi \over 4} {g N  \over \nn} \; .
\ee
We can satisfy (\ref{Lcond}) for integer $K$ provided that this quantity is much less than $N/2$, so we have the additional condition
\be
\nn \gg g \; .
\ee

So far, we have assumed that $k = k^{(0)}$. The actual value of $k$ corresponding to our chosen $L_A$s and $K$ is determined by
\be
{k \over \sqrt{g}} \nn + \sqrt{g} \sum_A F_k^{-1}(L_A) - N = 0
\ee
We need to check that for this actual value, $|k/\sqrt{g} \nn - N/2| \ll N$ so that (\ref{condB}) is still satisfied. Since 
\be
|{k^{(0)} \over \sqrt{g}} \nn + \sqrt{g} \sum_A F_{k^{(0)}}^{-1}(L_A) - N| \ll N
\ee
we know that the function 
\be
G(k) = {k \over \sqrt{g}} \nn + \sqrt{g} \sum_A F_k^{-1}(L_A)
\ee
varies by an amount much less than $N$ as $k$ is varied from $k^{(0)}$ to its actual value. This gives
\be
\delta k \ll {N \over G'(k)} \: ,
\ee
so $k \nn / \sqrt{g}$ will change by an amount much less than $N$ provided that the right side here is less than $ \sqrt{g} N / \nn$, or $G'(k) > \nn / \sqrt{g}$. This is clearly true, since the $k$ derivative of the first term in $G$ is $\nn / \sqrt{g}$ and the $k$ derivative of the second term is positive. 

To summarize, given $N$ and $g$, the following procedure will yield a set of linking numbers that satisfy our conditions:
\begin{itemize}
    \item Choose some $\nn$ satisfying $\nn \gg \sqrt{g N}$ and $\nn \gg g$ and $\nd$ satisfying (\ref{ndcond}).
    \item Choose a set $\{L_A\}$ of $\nd$ $L$s such that (\ref{L0cond}) and (\ref{Lcond}) are satisfied and 
    \be
    K = N_{D5} + {N - \sum_A L_A \over \nn}
    \ee
    is an integer. This will be possible provided the conditions on $\nn$ and $\nd$ are satisfied.
\item Once the linking numbers are fixed in this way, the precise $k$ and $l_A$ are determined by the procedure described at the beginning of this subsection.
\end{itemize}
For this more general class of SCFTs, the corresponding quiver gauge theory will have fundamental matter distributed among the nodes of the quiver, with the number of distinct $L_A$s determining the number of nodes with fundamental matter. 

If we require that $l_A < \epsilon \sqrt{N}$ to satisfy (\ref{L0cond}), we get
\be
L_{MAX} \approx F_{k^{(0)}}(\epsilon \sqrt{N}) = \epsilon \sqrt{gN} + {2 \over \pi} \nn \arctan \left({2 \epsilon \nn \over \sqrt{gN}}\right)
\ee
If $\nn \gg \sqrt{gN} / \epsilon$, we get $L_{MAX} \approx \nn$. As there are $\nn$ nodes in the quiver, it seems possible in some cases to have matter uniformly distributed throughout the quiver, with order one fundamentals per node.  

\section{Microscopic wedge holography} 
 
In this section we describe a generalization of the previous construction in which we have two ETW branes bounding an arbitrarily large wedge $\Theta \in (-\Theta_*,\Theta_*)$ of $AdS^5 \times S^5$. In this case, only a $\mathbb{R}^{2,1}$ of the original asymptotic region $\mathbb{R}^{3,1}$ of $AdS^5 \times S^5$ remains, and the dual theory is a three dimensional SCFT.
 
\subsection{A 3D dual to an arbitrarily large wedge of $AdS^5 \times S^5$.} \label{sec:3dwedge}

\begin{figure}
    \centering
    \includegraphics[height=2.5cm]{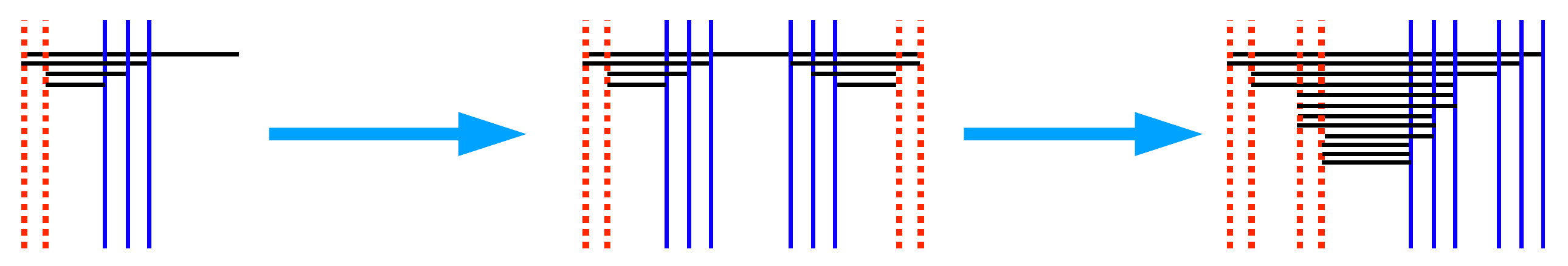}
    \caption{Illustration of procedure used to define families of solutions realizing arbitrarily large wedges of $AdS^{5} \times S^{5}$; here, D3-branes are black, D5-branes are blue, and NS5-branes are red. To pass from the second to the third configuration, we rearrange the fivebranes so that all NS5-branes are to the left of all D5-branes, while D3-branes between these fivebranes are created or annihilated to maintain fixed linking numbers. The third configuration is convenient for defining the quantities $N_{3}^{(A)}$, $\hat{N}_{3}^{(B)}$ in (\ref{eq:double}): they represent the net number of D3-branes ending on branes in the $A^{\textnormal{th}}$ D5-brane stack or the $B^{\textnormal{th}}$ NS5-brane stack respectively. }
    \label{fig:doubling}
\end{figure}

We have seen that for an appropriate choice of 3D SCFT coupled to ${\cal N}=4$ SYM theory on a half space, the ETW brane region of the dual geometry can be pushed to a Poincar\'e angle that is arbitrarily close to $-\pi/2$. We next consider the situation where we introduce another such boundary parallel to the first so that the ${\cal N}=4$ theory now lives on a strip. We can choose this second boundary SCFT to preserve the same set of supersymmetries as the first one. The brane construction of this SCFT involves the same set of branes as for the first SCFT, with the same orientations, but arranged in the opposite order in the spatial direction in which the D3-branes have a boundary;\footnote{More generally, we could consider two different SCFTs which nevertheless preserve the same supersymmetries.} see Figure \ref{fig:doubling}.

We expect the dual of this theory to have two ETW branes, bounding a wedge of $AdS^5 \times S^5$ whose asymptotic region has the geometry $\mathbb{R}^{2,1}$ times an interval. The solutions of \cite{DHoker:2007zhm, DHoker:2007hhe, Aharony:2011yc, Assel:2011xz} are not general enough to describe this, since they correspond to theories with a 3D superconformal symmetry, while the interval in our construction introduces a scale. However, we expect that the IR limit of the theory on a strip will be a certain superconformal theory; this is the theory whose brane construction combines that of the original BCFT with that of the second SCFT, so that the initial semi-infinite D3-branes now connect the brane configurations describing the two SCFTs. The gravity dual of this IR SCFT is wedge of $AdS^5 \times S^5$ with two ETW branes. This wedge geometry can be described explicitly as particular cases of the solutions in \cite{DHoker:2007zhm, DHoker:2007hhe, Aharony:2011yc, Assel:2011xz} and were considered previously in \cite{Assel:2011xz, Bachas:2017rch}. These geometries are microscopic realizations of the ``wedge holography'' discussed in \cite{Akal:2020wfl}.

The new element in our work is that we can, by the choices described in the previous section, arrange for the wedge of $AdS^5 \times S^5$ between the ETW branes to be arbitrarily large.

To verify this, we note that making the change of coordinates $z = r_0 e^w = r_0 e^{x+iy}$ so that the positive quadrant is mapped to the strip $0 \le \Im(w) \le \pi/2$, the single boundary geometries correspond to harmonic functions 
\beas
h_1 & = & \frac{\pi \ell_{s}^{2}}{2 \sqrt{g}} r_{0} e^{x} \cos y + \frac{\ell_{s}^{2}}{4 \sqrt{g}} \sum_{A} c_{A} \ln \left( \frac{\cosh(x + \alpha_{A}) + \cos(y)}{\cosh(x + \alpha_{A}) - \cos(y)} \right) \cr
h_2 & = & \frac{\pi \ell_{s}^{2} \sqrt{g}}{2} r_{0} e^{x} \sin y + \frac{\ell_{s}^{2} \sqrt{g}}{4 } \sum_{B} d_{B} \ln \left( \frac{\cosh(x + \beta_{B}) + \sin(y)}{\cosh(x + \beta_{B}) - \sin(y)} \right) \: ,
\eeas
where we have defined $\alpha_A = -\ln(l_A/r_0)$ and $\beta_A = -\ln(k_A/r_0)$. 

The pole of $h_1$ at $-\alpha$ and the pole of $h_2$ at $i \pi/2 - \beta$ lie at large negative values of $x$ for the single-pole cases of interest. The corresponding solution with two ETW branes is given by
\beas
h_1 = \frac{\ell_{s}^{2}}{4} \sum_{a=1}^{2} N_{5}^{(a)} \ln \left( \frac{\cosh(x - \delta_{a}) + \cos(y)}{\cosh(x - \delta_{a}) - \cos(y)} \right) \cr
h_2 = \frac{\ell_{s}^{2}}{4} \sum_{b=1}^{2} \hat{N}_{5}^{(b)} \ln \left( \frac{\cosh(x - \hat{\delta}_{b}) + \sin(y)}{\cosh(x  - \hat{\delta}_{b}) - \sin(y)} \right) \: ,
\eeas
where $N_{5}^{(1)} = N_{5}^{(2)} = N_{D5}$ and $\hat{N}_{5}^{(1)} = \hat{N}_{5}^{(2)} = N_{NS5}$ are the number of D5-branes and NS5-branes in the initial boundary condition, and
now we have poles of $h_1$ at $\delta_{1/2}$ and of $h_{2}$ at $i \pi/2 + \hat{\delta}_{1/2}$ whose leading order behaviour is given by
\begin{equation}
    \delta_{1} \sim - \delta_{2} \sim \alpha \: , \qquad \hat{\delta}_{1} \sim - \hat{\delta}_{2} \sim \beta \: .
\end{equation} 
Solutions corresponding to more general 3D SCFTs are obtained by allowing the poles to be at more general locations.

To demonstrate this claim, we will proceed by analyzing the D-brane constructions for these theories. We must first revisit the families of boundary conditions from the previous section, choosing for convenience a string coupling $g$ in the boundary case to be of the form $g = m \cot \left( \frac{\pi}{2} \frac{a}{b} \right)$,
as we have done above, and defining the parameters $(N_{D5}, N_{NS5}, L, K)$ using (\ref{eq:generalseq}). As in \cite{Assel:2011xz}, when we pass to the dual of the 3D theory, we may consistently set $g=1$ (while the dilaton is left arbitrary). 

The doubled theory is described in the language of \cite{Assel:2011xz} by parameters\footnote{Our notation is actually slightly different from that of \cite{Assel:2011xz}: the $N_{3}^{(i)}$ and $\hat{N}_{3}^{(i)}$ are both defined to be positive quantities, and differ from the conventions of that reference by factors of $N_{5}^{(i)}$ and $\hat{N}_{5}^{(i)}$ respectively.}
\begin{equation} \label{eq:double}
    \begin{split}
        N_{5}^{(1)} & = N_{5}^{(2)} = N_{D5} \: , \\  \hat{N}_{5}^{(1)} & = \hat{N}_{5}^{(2)} = N_{NS5} \: , \\
        N_{3}^{(1)} & = 2 N_{NS5} - L \: , \quad N_{3}^{(2)} = L \: , \\
        \hat{N}_{3}^{(1)} & = K \: , \quad \hat{N}_{3}^{(2)} = 2 N_{D5} - K \: ,
    \end{split}
\end{equation}
where the supergravity parameters $\delta_{a}, \hat{\delta}_{b}$ are related to the D3-brane charges by
\begin{equation} \label{eq:3dquantization}
    \begin{split}
        N_{3}^{(a)} & = \frac{2}{\pi} \sum_{b=1}^{2} \hat{N}_{5}^{(b)} \tan^{-1} \left( e^{ \delta_{a} - \hat{\delta}_{b}} \right) \\
        \hat{N}_{3}^{(b)} & = \frac{2}{\pi} \sum_{a=1}^{2} N_{5}^{(a)} \tan^{-1} \left( e^{ \delta_{a} - \hat{\delta}_{b}} \right) \: .
    \end{split}
\end{equation}

These latter equations yield at leading order in $n$ 
\begin{equation}
    \begin{split}
        e^{\delta_{1} - \hat{\delta}_{1}} = \frac{g}{m} \: , \quad e^{\delta_{1} - \hat{\delta}_{2}}  = \frac{4 m b^{2} n^{2}}{\pi N} \: , \quad e^{\delta_{2} - \hat{\delta}_{1}} = \frac{\pi N}{4 m b^{2} n^{2}} \: , \quad e^{\delta_{2} - \hat{\delta}_{2}}  = \frac{m}{g} \: ,
    \end{split}
\end{equation}
so that without loss of generality we may take leading order behaviour
\begin{equation}
   e^{\delta_{1}} = \frac{g}{m} e^{\hat{\delta}_{1}} = e^{- \delta_{2}} = \frac{g}{m} e^{- \hat{\delta}_{2}} = \frac{2 \sqrt{g} b n}{\sqrt{\pi N}} 
    \: .
\end{equation}
Comparing with the supergravity parameters from the boundary case
\begin{equation}
    \frac{l}{r_{0}} \sim \frac{\sqrt{\pi N}}{2 \sqrt{g} b n} \: , \qquad \frac{k}{r_{0}} \sim \frac{\sqrt{g \pi N}}{2 m b n} \: ,
\end{equation}
we find the leading behaviour of the poles $\delta_{1/2}$ and $\hat{\delta}_{1/2}$ mentioned above.

One can consider $h_{1}, h_{2}$ at leading order, and show that they give rise to an AdS$_{5} \times S^{5}$ region when $|x| \ll \ln n$. Indeed, we find in this region
\begin{equation}
\begin{split}
    h_{1} & \sim  \frac{ L^{2}}{2 \sqrt{g}}  \cosh x \cos y  \: , \\
    h_{2} & \sim \frac{\sqrt{g} L^{2}}{2} \cosh x \sin y \: ,
\end{split}
\end{equation}
where $L^{2} = \sqrt{4 \pi N} \ell_{s}^{2}$. 
We recognize these as corresponding to pure $AdS^{5} \times S^{5}$. 
As $n$ is increased, the curvature scale of the $AdS^{5} \times S^{5}$ region approaches a constant value, while the size of this region increases.

\begin{figure}[h]
\centering
\begin{subfigure}{.5\textwidth}
  \centering
  \includegraphics[height=6.5cm]{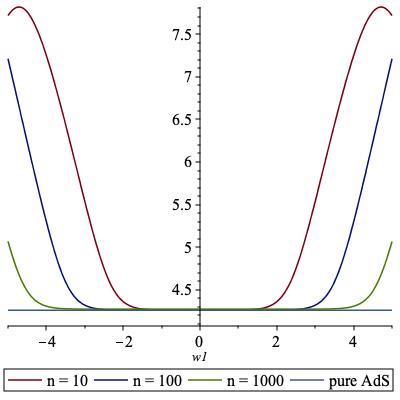}
  \caption{$\ln f_{1}^{2}$ versus $w_{1}$}
  \label{fig:f1_2_3D}
\end{subfigure}%
\begin{subfigure}{.5\textwidth}
  \centering
  \includegraphics[height=6.5cm]{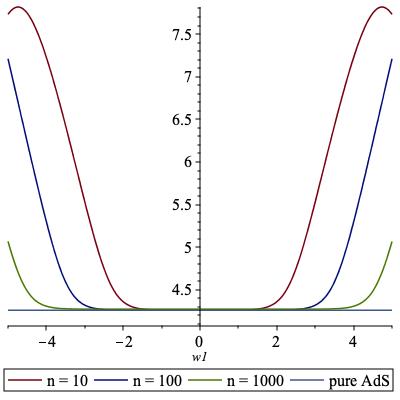}
  \caption{$\ln f_{2}^{2}$ versus $w_{1}$}
  \label{fig:f2_2_3D}
\end{subfigure} 
\\
\begin{subfigure}{.49\textwidth}
  \centering
  \includegraphics[height=6.5cm]{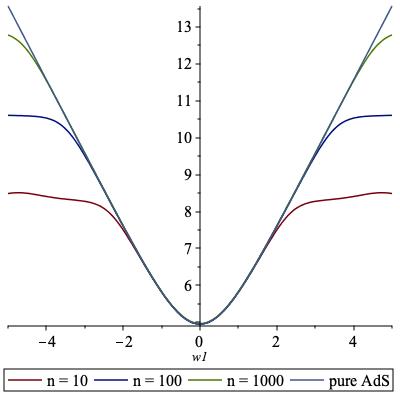}
  \caption{$\ln f_{4}^{2}$ versus $w_{1}$}
  \label{fig:f4_2_3D}
\end{subfigure}
\begin{subfigure}{.49\textwidth}
  \centering
  \includegraphics[height=6.5cm]{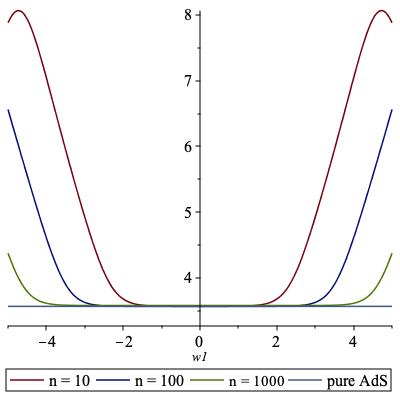}
  \caption{$\ln \rho^{2}$ versus $w_{1}$}
  \label{fig:rho_2_3D}
\end{subfigure}
\caption{In these figures, we are taking $g=1, \ell_{\textnormal{s}} = 2, N = 100$. The metric functions shown are for $N_{5} = 2 n N, \hat{N}_{5} = 2 (n N + 2), \hat{N}_{3}^{(1)} = \frac{N}{2} (n+1)$ with the values of $n$ given, while the metric functions shown in light blue are for pure $AdS^{5} \times S^{5}$ (with $L_{\textnormal{AdS}}$ fixed by $N$). We are displaying the metric functions with respect to complex coordinates $(w, \bar{w}) = (w_{1} + i w_{2}, w_{1} - i w_{2}) = \left( \ln \left( \frac{r}{r_{0}} e^{i \theta}\right), \ln \left( \frac{r}{r_{0}} e^{- i \theta} \right) \right)$, and setting $\theta = \frac{\pi}{4}$ in the figures. Note that the Jacobian of this coordinate change modifies $\rho^{2}$ from the expression provided.}
\label{fig:metricfunctions_3D}
\end{figure}

In Figure \ref{fig:metricfunctions_3D}, we show the metric functions for such solutions (as well as those of $AdS^{5} \times S^{5}$ for comparison) in the vicinity of the locally $AdS^5 \times S^5$ bridge between the two ETW branes, for various increasing values of $n$. We see that for increasing $n$, the bridge connecting the two ETW brane regions corresponds to an increasingly large wedge of $AdS^5 \times S^5$.

\subsection{Multi-wedge geometries}

We have given a specific class of constructions describing arbitrarily large wedges of $AdS^5 \times S^5$ as the dual of a 3D SCFT. For concreteness, we focused on the case obtained by doubling of a brane configuration considered earlier in the context of boundary conditions for the 4D $\mathcal{N}=4$ theory. More generally, we may consider 3D SCFTs which descend from linear quivers arising from ``gluing" together several large sub-quivers of the type discussed in Section \ref{sec:fieldtheory} by coupling the first and last nodes of consecutive sub-quivers with bifundamental matter to some additional $U(m_{A})$ nodes with small $m_{A}$. 
This procedure is in the spirit of the ``quantum gate" solutions described by Bachas and Lavdas in \cite{Bachas:2017rch}, but the result here is a spacetime description involving multiple wedges of $AdS^{5} \times S^{5}$ separated by interface branes. 


This ``multi-wedge" construction suggests further generalizations for holographic theories realizing the same $OSp(4|4)$ symmetry as the 3D SCFTs, including the $OSp(4|4)$-preserving BCFTs and 3D SCFTs descending from circular quiver gauge theories. In the former case, the holographic description involves a large $AdS^{5} \times S^{5}$ region in the vicinity of the asymptotic boundary, but this region is connected to an additional multi-wedge region by an interface brane. In the latter case, we again obtain a multi-wedge geometry whose boundary is only a $\mathbb{R}^{2, 1}$ subset of the asymptotic $\mathbb{R}^{3,1}$ of $AdS^{5} \times S^{5}$, but in this case, the first and last $AdS^{5} \times S^{5}$ wedges are connected by another interface brane, so that we have non-contractible loops in the internal space which traverse all of the wedges. We leave a more detailed analysis of multi-wedge solutions to Appendix \ref{app:multi}.

\section{Discussion}

We have provided a number of microscopic constructions of 4D BCFTs enjoying a holographic description with an arbitrarily large $AdS^{5} \times S^{5}$ region terminating on an ETW brane, as well as 3D SCFTs which correspond to an arbitrarily large $AdS^{5} \times S^{5}$ wedge. 
While the possibility of realizing similar features by considering limits of the supergravity solutions provided in \cite{Aharony:2011yc, Assel:2011xz, Assel:2012cj} has been discussed previously (e.g. in \cite{Assel:2011xz, Assel:2012cj, Bachas:2017rch, Bachas:2018zmb}), we have provided an important check that the required limits can indeed be realized in string theory, where the various charges are subject to quantization requirements, and we have characterized the appropriate boundary conditions explicitly in terms of the corresponding field theory data.

The simplest such BCFT boundary conditions arise in string theory from a single stack of $N_{D5}$ D5-branes and $N_{NS5}$ NS5-branes; choosing $N_{D5}, N_{NS5}$ sufficiently large with $g N_{D5} / N_{NS5} = O(1)$ ensures a large $AdS^{5} \times S^{5}$ region, and a judicious choice of these parameters and the linking numbers $L, K$ can make this region arbitrarily large. While these ``single-pole" boundary conditions are especially easy to analyze, we have indicated several generalizations involving multiple fivebrane throats in the ETW brane region, including small perturbations to the single-pole boundary conditions, boundary conditions which redistribute the fundamental matter throughout the defining quiver diagram,
and boundary conditions involving extended quivers which give rise to ``multi-wedge" duals. By invoking similar D-brane constructions to generate supersymmetric boundary conditions for the 4D $\mathcal{N}=4$ SYM theory or 3D SCFTs describing the IR physics of linear or circular quiver gauge theories, we are able to produce holographic duals for these theories in type IIB supergravity that possess similar local features, including one or more $AdS^{5} \times S^{5}$ wedges. This suggests a precise sense in which the physics of these degrees of freedom can be associated to the wedge. In all of our examples, such wedges are necessarily accompanied by a large ETW brane region. 

There are a number of further directions which remain interesting to explore. While we have studied a large class of solutions with large $AdS^{5} \times S^{5}$ regions, it would be desirable to provide a general characterization of theories which possess this feature. It is also interesting to understand if there is a relationship between our work and the  ``dimensional (de)construction" story \cite{ArkaniHamed:2001ca, ArkaniHamed:2001ie}. In this context, it is shown that certain quiver gauge theories may admit a low energy effective description with emergent extra dimensions; for example, this may occur in superconformal theories moved onto the Higgs branch, with the spectrum of massive vectors obtained via the Higgs mechanism organizing precisely into the Kaluza-Klein modes of the higher dimensional theory. Our results also suggest a relationship between 3D and 4D supersymmetric theories, in the sense
that the physics of large wedges of $AdS^{5} \times S^{5}$ can either be described by degrees of freedom in the 4D $\mathcal{N}=4$ SYM theory or in a suitably chosen 3D SCFT capturing the low energy behaviour of a quiver gauge theory.

\section*{Acknowledgments}

This research is supported in part by the Natural Sciences and Engineering Research Council of Canada and by the Simons Foundation though a Simons Investigator award and the ``It From Qubit'' Collaboration grant.

\appendix

\section{Size of the internal space in the ETW brane region} \label{app:internal}

The fact that the compact spherical directions in the ``bag" or ``ETW brane" region of the geometries of interest cannot be suppressed relative to the $AdS^{4}$ scale has already been noted by Bachas and Lavdas in \cite{Bachas:2018zmb} (following previous related comments by Bachas and Estes in \cite{Bachas:2011xa}). As remarked by these authors, this property is related to the issue of scale separation in the context of flux compactifications
(see e.g. \cite{Polchinski:2009ch, Tsimpis:2012tu, Gautason:2015tig}). More generally, it is a broad prediction that in holographic theories with supersymmetry, the R-symmetry is geometrized at the AdS scale (see e.g.
\cite{Alday:2019qrf}). For the sake of completeness, we will here provide a direct argument for these assertions in the context of the supergravity solutions considered in this note, based on the formulation of the reduced BPS equations by D'Hoker, Estes, and Gutperle in \cite{DHoker:2007zhm, DHoker:2007hhe}.
Our conclusions will apply to the solutions dual to the 3D $\mathcal{N}=4$ SCFTs of Gaiotto-Witten \cite{Gaiotto:2008sa, Gaiotto:2008ak}, first studied in \cite{Assel:2011xz}, as well as the boundary and interface solutions studied in \cite{Aharony:2011yc}.

Our goal is to show that it is not possible to simultaneously have $f_{1}^{2} / f_{4}^{2} \ll 1$ and $f_{2}^{2} / f_{4}^{2} \ll 1$ in \textit{any} region of the spacetime unless that region is locally $AdS^{5} \times S^{5}$; the conclusion is therefore that at least one of the $S^{2}$ factors of the internal space remains large relative to the $AdS^{4}$ scale in the ETW brane region.

In the following, we will be relying on the conventions of \cite{DHoker:2007zhm}, introducing only the ingredients necessary. 
We may write the complex axion/dilaton $P$ and connection $Q$ one-forms as
\begin{equation}
    P = p_{a} e^{a} \: , \qquad Q = q_{a} e^{a} \: ,
\end{equation}
and the anti-symmetric five-form and three-form tensors $F_{(5)}$ and $G$ as
\begin{equation}
    F_{(5)} = f_{a} \left( - e^{0123a} + \varepsilon^{a}_{\: b} e^{4567b} \right) \: , \quad G = g_{a} e^{45a} + i h_{a} e^{67a} \: ,
\end{equation}
where the $e$ are wedge products of the appropriate vielbeins; 
the indices $a, b$ are summed over the Riemann surface $\Sigma$ directions. It is demonstrated in \cite{DHoker:2007zhm} that, for solutions with 16 supersymmetries, one can always apply an $SU(1, 1)$ S-duality transformation to a frame where the axion field vanishes and the dilaton is real; this corresponds to the reality conditions
\begin{equation}
    \bar{p}_{a} = p_{a} \: , \quad \bar{g}_{a} = g_{a} \: , \quad \bar{h}_{a} = h_{a} \: , \quad q_{a} = 0 \: .
\end{equation}

The metric functions $f_{1}, f_{2}, f_{4}$ may be expressed in terms of a (Grassmann-even) spinor degree of freedom (equation (6.18) of \cite{DHoker:2007zhm})
\begin{equation}
    \xi = \begin{pmatrix}
        \alpha \\ \beta
    \end{pmatrix}
    \: , \qquad 
    \xi^{*} = \begin{pmatrix}
        \bar{\alpha} \\ \bar{\beta}
    \end{pmatrix}
    \: , \qquad \alpha, \beta \in \mathbb{C} \: ,
\end{equation}
in terms of which we have
(equation (6.26) of \cite{DHoker:2007zhm})
\begin{equation}
    \begin{split}
        f_{4} & = \xi^{\dagger} \xi = \alpha \bar{\alpha} + \beta \bar{\beta} \\
        f_{1} & = - \nu \xi^{\dagger} \sigma^{1} \xi = - \nu \left( \alpha \bar{\beta} + \beta \bar{\alpha} \right) \\
        f_{2} & = - \xi^{\dagger} \sigma^{2} \xi = i \left( \beta \bar{\alpha} - \alpha \bar{\beta} \right) \: ,
    \end{split}
\end{equation}
where $\nu \in \{ \pm 1\}$ (the sign will be irrelevant when we compare ratios of metric functions $f_{1}^{2}, f_{2}^{2}$ and $f_{4}^{2}$). 

Suppose there is some neighbourhood of a point $(w, \bar{w})$ in the interior of our geometry where $f_{1}^{2} / f_{4}^{2} \ll 1$ and $f_{2}^{2} / f_{4}^{2} \ll 1$; we will restrict to considering this neighbourhood for the remainder of the subsection. In this case, we must have either $|\alpha| \ll |\beta|$ or $| \beta| \ll | \alpha |$ throughout the neighbourhood. Indeed, using polar coordinates
\begin{equation}
    \alpha = a e^{i \theta_{1}} \: , \qquad \beta = b e^{i \theta_{2}} \: , 
\end{equation}
we have
\begin{equation}
    \begin{split}
        \Bigg| \frac{f_{1}}{f_{4}} \Bigg| & = \frac{2 a b}{a^{2} + b^{2}} \big| \cos \left( \theta_{1} - \theta_{2} \right) \big| \\
        \Bigg| \frac{f_{2}}{f_{4}} \Bigg| & = \frac{2 a b}{a^{2} + b^{2}} \big| \sin \left( \theta_{1} - \theta_{2} \right) \big| \: ,
    \end{split}
\end{equation}
and since
\begin{equation}
    \min_{\theta_{1}, \theta_{2}} \max \{ \big| \cos \left( \theta_{1} - \theta_{2} \right) \big| , \big| \sin \left( \theta_{1} - \theta_{2} \right) \big| \} = \frac{1}{\sqrt{2}} \: ,
\end{equation}
we must have $\frac{2 a b}{a^{2} + b^{2}} \ll 1$, which requires $a \ll b$ or $b \ll a$. 

On the other hand, the dilatino BPS equation (equation (6.28) of \cite{DHoker:2007zhm}) gives
\begin{equation}
    4 p_{z} \alpha + \left( g_{z} - i h_{z} \right) \beta = 0 \: , \quad 4 p_{z} \beta - \left( g_{z} + i h_{z} \right) \alpha = 0 \: ,
\end{equation}
with $z, \bar{z}$ frame indices. These two equations together imply either that $p_{z} = g_{z} = h_{z} = 0$ or
\begin{equation}
    \Bigg| \frac{\alpha}{\beta} \Bigg| = \Bigg| \frac{4 p_{z}}{g_{z} + i h_{z}} \Bigg| = \Bigg| \frac{4 p_{z}}{g_{z} - i h_{z}} \Bigg| = \Bigg| \frac{\beta}{\alpha}\Bigg| \: ,
\end{equation}
with the latter contradicting the conclusion that $|\alpha| \ll |\beta|$ or $| \beta | \ll | \alpha |$. We therefore must have that the special condition $p_{z} = g_{z} = h_{z} = 0$ holds throughout the neighbourhood we are considering.\footnote{Note that we could have avoided this condition by requiring that \textit{one of} $f_{1}^{2}/f_{4}^{2}$ or $f_{2}^{2} / f_{4}^{2}$ \textit{but not both} was small; in this case, we would not necessarily require that $|\alpha| \ll |\beta|$ or $|\beta| \ll |\alpha|$, but could instead have that $\alpha \bar{\beta}$ was almost pure real or pure imaginary. } 
But as shown in Section 6.9 of \cite{DHoker:2007zhm}, that condition alone necessarily implies that the geometry is pure $AdS^{5} \times S^{5}$, with (subject to a particular choice of normalization)
\begin{equation}
    \alpha = e^{- \nu w / 2} \: , \qquad \beta = i e^{\nu w / 2} \: ,
\end{equation}
and metric functions
\begin{equation}
    f_{1} = 2 \sin y \: , \qquad f_{2} = - 2 \cos y \: , \qquad f_{4} = 2 \cosh x \: ,
\end{equation}
where $w = x + i y$ is a complex coordinate on the strip $\Sigma$. 
(We should note that the argument provided applies to the case where $p_{z}, g_{z}, h_{z}$ are presumed to vanish everywhere, but the nature of the argument is local, and can be repeated to demonstrate that the geometry within the neighbourhood we are considering must be $AdS^{5} \times S^{5}$.)
In particular, this can be consistent with our assumption $|\alpha| \ll |\beta|$ or $|\beta| \ll |\alpha|$ near the asymptotic boundary $x \rightarrow \pm \infty$, where the metric function $f_{4}^{2}$ diverges. 
We have therefore shown that the only case in which one can simultaneously have $f_{1}^{2} / f_{4}^{2} \ll 1$ and $f_{2}^{2} / f_{4}^{2} \ll 1$ is when the geometry is locally $AdS^{5} \times S^{5}$; as a corollary, we clearly cannot have the scale of the internal $S^{2}$ dimensions be small compared to the curvature scale of the non-compact dimensions.

\section{Justification of condition (\ref{deltacond})} \label{app:cldk_justification}


In general, the \textit{region I} introduced in Section \ref{sec:obtaining} is only asymptotically $AdS^{5} \times S^{5}$, and may deviate from pure $AdS^{5} \times S^{5}$ significantly before the $O(l_{A}/r)$, $O(k_{B}/r)$ corrections become large. For example, considering the large-$r$ asymptotics of the metric functions for our general solution, we find
\begin{equation}
    \begin{split}
        \rho^{2} & = \frac{L_{\textnormal{AdS}}^{2}}{4} \frac{1}{r^{2}} \left[ 1 - \frac{1}{r^{2}} \left( 2 \cos^{2} \theta - 1 \right) \left( \frac{1}{\pi} \sum_{A} c_{A} l_{A} \left(  1 - \frac{l_{A}^{2}}{r_{0}^{2}} \right)
        - \frac{1}{\pi} \sum_{B} d_{B} k_{B} \left( 1 - \frac{k_{B}^{2}}{r_{0}^{2}} \right)
        \right)  + o(r^{-2}) \right] \\
        f_{1}^{2} & = L_{\textnormal{AdS}}^{2} \cos^{2} \theta \Bigg[ 1 + \frac{1}{r^{2}} \Bigg( \frac{1}{\pi} \sum_{A} c_{A} l_{A} \left( \left( 2 \cos^{2} \theta + 1 \right) + \frac{l_{A}^{2}}{r_{0}^{2}} \left( 2 \cos^{2} \theta - 1 \right) \right) \\
        & \qquad \qquad \qquad - \frac{1}{\pi} \sum_{B} d_{B} k_{B} \left( \left( 2 \cos^{2} \theta + 1 \right) + \frac{k_{B}^{2}}{r_{0}^{2}} \left( 2 \cos^{2} \theta - 1 \right) \right) \Bigg) + o(r^{-2}) \Bigg] \\
        f_{2}^{2} & = L_{\textnormal{AdS}}^{2} \sin^{2} \theta \Bigg[ 1 + \frac{1}{r^{2}} \Bigg( \frac{1}{\pi} \sum_{A} c_{A} l_{A} \left( \left( 2 \cos^{2} \theta - 3 \right) + \frac{l_{A}^{2}}{r_{0}^{2}} \left( 2 \cos^{2} \theta - 1 \right) \right) \\
        & \qquad \qquad \qquad - \frac{1}{\pi} \sum_{B} d_{B} k_{B} \left( \left( 2 \cos^{2} \theta - 3 \right) + \frac{k_{B}^{2}}{r_{0}^{2}} \left( 2 \cos^{2} \theta - 1 \right) \right) \Bigg) + o(r^{-2}) \Bigg]  \\
        f_{4}^{2} & = \frac{L_{\textnormal{AdS}}^{2} (r^{2} + r_{0}^{2})^{2}}{4 r_{0}^{2} r^{2}}
         \\
         & \qquad - \frac{L_{\textnormal{AdS}}^{2} r^{2}}{4 r_{0}^{2}} \left[ \frac{1}{r^{2}} \left( 2 \cos^{2} \theta - 1 \right) \left( \frac{1}{\pi} \sum_{A} c_{A} l_{A} \left( 1 + \frac{l_{A}^{2}}{r_{0}^{2}} \right) - \frac{1}{\pi} \sum_{B} d_{B} k_{B} \left( 1 + \frac{k_{B}^{2}}{r_{0}^{2}} \right) \right) + o(r^{-2}) \right]  \: .
    \end{split}
\end{equation}
Evidently, if we would like the terms subleading in large $r$ to be suppressed for any $r \ll r_{0}$, then in addition to (\ref{klcond}), we require
\begin{equation} \label{eq:purecond}
     \big| \sum_{A} c_{A} l_{A} - \sum_{B} d_{B} k_{B} \big| \ll r_*^{2} \: .
\end{equation}
We claim that conditions (\ref{klcond}) and (\ref{eq:purecond}) are sufficient to ensure a large region of approximately pure $AdS^{5} \times S^{5}$.

To further motivate this fact, let us fix $N$ from the beginning, and recall that $r_{0}^{2} \equiv \frac{N}{\pi}$.
Suppose we would like to have a geometry well-approximated by pure $AdS^{5} \times S^{5}$ down to some radial coordinate $r_* \ll r_{0}$. 
Our approach will be to write down the metric functions in the limit $\frac{l_{A}}{r}, \frac{k_{B}}{r} \rightarrow 0$ with $\sum_{A} c_{A} l_{A}$ and $\sum_{B} d_{B} k_{B}$ held fixed, and to understand how these functions depend on the quantity $\Big| \sum_{A} c_{A} l_{A} - \sum_{B} d_{B} k_{B}\Big|$. In particular, letting 
\begin{equation}
    \sum_{A} c_{A} l_{A} = \frac{\pi}{2} r_{0}^{2} \left( 1 + \varepsilon \right) \: , \qquad \sum_{B} d_{B} k_{B} = \frac{\pi}{2} r_{0}^{2} \left( 1 - \varepsilon \right) \: ,
\end{equation}
we find that when $\frac{l_{A}}{r}, \frac{k_{B}}{r} \rightarrow 0$ in a way that keeps $N$ and $\varepsilon$ fixed, we have
\begin{equation}
    \begin{split}
        \left( \frac{\pi}{2} \right)^{-1} h_{1}(r, \theta) & = r \cos \theta + \frac{r_{0}^{2} \cos \theta}{r} \left( 1 + \varepsilon \right) \\
        \left( \frac{\pi}{2} \right)^{-1} h_{2}(r, \theta) & = r \sin \theta + \frac{r_{0}^{2} \sin \theta}{r} \left( 1- \varepsilon \right) \\
        \left( \frac{\pi}{2} \right)^{-2} W(r, \theta) & = - \frac{2 r_{0}^{2} \sin \theta \cos \theta}{r^{2}} 
    \end{split}
\end{equation}
and
\begin{equation}
    \begin{split}
        \left( \frac{\pi}{2} \right)^{-4} N_{1}(r, \theta) & = \frac{\sin \theta \cos \theta}{2} \left( r^{2} + r_{0}^{2} (1 + \varepsilon) \right) \Big[ 1 + \frac{ r_{0}^{2}}{r^{2}} \left( 3 + \varepsilon (1 - 4 \cos^{2} \theta) \right) \\
        & \qquad + \frac{r_{0}^{4}}{r^{4}} \left( 1 + \varepsilon \right) \left( 3 - \varepsilon (1 - 4 \cos^{2} \theta) \right)  + \frac{r_{0}^{6}}{r^{6}} \left( 1 + \varepsilon \right)^{2} \left( 1 - \varepsilon \right) \Big] 
    \end{split}
\end{equation}
\begin{equation}
    \begin{split}
        \left( \frac{\pi}{2} \right)^{-4} N_{2}(r, \theta) & = \frac{\sin \theta \cos \theta}{2} \left( r^{2} + r_{0}^{2} (1 - \varepsilon) \right) \Big[ 1 + \frac{ r_{0}^{2}}{r^{2}} \left( 3 + \varepsilon (3 - 4 \cos^{2} \theta) \right) \\
        & \qquad + \frac{r_{0}^{4}}{r^{4}} \left( 1 - \varepsilon \right) \left( 3 - \varepsilon (3 - 4 \cos^{2} \theta) \right)  + \frac{r_{0}^{6}}{r^{6}} \left( 1 - \varepsilon \right)^{2} \left( 1 + \varepsilon \right) \Big] \: . 
    \end{split}
\end{equation}
We then find the metric function
\begin{equation}
    \begin{split}
        \rho^{2}(r, \theta) & = \frac{L^2}{4 r^{2}} \left( 1 + \frac{r_{0}^{2}}{r^{2}} \left( 1- \varepsilon \right) \right)^{-3/4} \left( 1 + \frac{r_{0}^{2}}{r^{2}} \left( 1 + \varepsilon \right) \right)^{-3/4} \\
        & \quad \Bigg[ \left( \left( 1 + \frac{r_{0}^{2}}{r^{2}} \left( 1- \varepsilon \right) \right) \left( 1 + \frac{r_{0}^{2}}{r^{2}} \left( 1 + \varepsilon \right) \right)^{2} - 4 \varepsilon \cos^{2} \theta \frac{r_{0}^{2}}{r^{2}} \left( 1 - \frac{r_{0}^{2}}{r^{2}} \left( 1 + \varepsilon \right) \right) \right) \\
        & \quad \times \left( 1 + \frac{r_{0}^{2}}{r^{2}} \left( 3 + \varepsilon (3 - 4 \cos^{2} \theta) \right) + \frac{r_{0}^{4}}{r^{4}} \left( 4 \varepsilon (1-\varepsilon) + 3 (1-\varepsilon)^{2} \right) + \frac{r_{0}^{6}}{r^{6}} (1-\varepsilon)^{2} (1+\varepsilon)
        \right)
        \Bigg]^{1/4} 
    \end{split}
\end{equation}
\begin{equation}
    \begin{split}
        f_{1}^{2}(r, \theta) & = L^{2} \cos^{2} \theta \left( 1 + \frac{r_{0}^{2}}{r^{2}} \left( 1- \varepsilon \right) \right)^{1/4} \left( 1 + \frac{r_{0}^{2}}{r^{2}} \left( 1 + \varepsilon \right) \right)^{5/4} \\
        & \quad \Bigg[ \left( \left( 1 + \frac{r_{0}^{2}}{r^{2}} \left( 1- \varepsilon \right) \right) \left( 1 + \frac{r_{0}^{2}}{r^{2}} \left( 1 + \varepsilon \right) \right)^{2} - 4 \varepsilon \cos^{2} \theta \frac{r_{0}^{2}}{r^{2}} \left( 1 - \frac{r_{0}^{2}}{r^{2}} \left( 1 + \varepsilon \right) \right) \right)^{-3} \\
        & \quad \times \left( 1 + \frac{r_{0}^{2}}{r^{2}} \left( 3 + \varepsilon (3 - 4 \cos^{2} \theta) \right) + \frac{r_{0}^{4}}{r^{4}} \left( 4 \varepsilon (1-\varepsilon) + 3 (1-\varepsilon)^{2} \right) + \frac{r_{0}^{6}}{r^{6}} (1-\varepsilon)^{2} (1+\varepsilon)
        \right)
        \Bigg]^{1/4} 
    \end{split}
\end{equation}
\begin{equation}
    \begin{split}
        f_{2}^{2}(r, \theta) & = L^{2} \sin^{2} \theta \left( 1 + \frac{r_{0}^{2}}{r^{2}} \left( 1- \varepsilon \right) \right)^{5/4} \left( 1 + \frac{r_{0}^{2}}{r^{2}} \left( 1 + \varepsilon \right) \right)^{1/4} \\
        & \quad \Bigg[ \left( \left( 1 + \frac{r_{0}^{2}}{r^{2}} \left( 1- \varepsilon \right) \right) \left( 1 + \frac{r_{0}^{2}}{r^{2}} \left( 1 + \varepsilon \right) \right)^{2} - 4 \varepsilon \cos^{2} \theta \frac{r_{0}^{2}}{r^{2}} \left( 1 - \frac{r_{0}^{2}}{r^{2}} \left( 1 + \varepsilon \right) \right) \right) \\
        & \quad \times \left( 1 + \frac{r_{0}^{2}}{r^{2}} \left( 3 + \varepsilon (3 - 4 \cos^{2} \theta) \right) + \frac{r_{0}^{4}}{r^{4}} \left( 4 \varepsilon (1-\varepsilon) + 3 (1-\varepsilon)^{2} \right) + \frac{r_{0}^{6}}{r^{6}} (1-\varepsilon)^{2} (1+\varepsilon)
        \right)^{-3}
        \Bigg]^{1/4} 
    \end{split}
\end{equation}
\begin{equation}
    \begin{split}
        f_{4}^{2}(r, \theta) & = \frac{L^{2} r^{2}}{4 r_{0}^{2}} \left( 1 + \frac{r_{0}^{2}}{r^{2}} \left( 1- \varepsilon \right) \right)^{1/4} \left( 1 + \frac{r_{0}^{2}}{r^{2}} \left( 1 + \varepsilon \right) \right)^{1/4} \\
        & \quad \Bigg[ \left( \left( 1 + \frac{r_{0}^{2}}{r^{2}} \left( 1- \varepsilon \right) \right) \left( 1 + \frac{r_{0}^{2}}{r^{2}} \left( 1 + \varepsilon \right) \right)^{2} - 4 \varepsilon \cos^{2} \theta \frac{r_{0}^{2}}{r^{2}} \left( 1 - \frac{r_{0}^{2}}{r^{2}} \left( 1 + \varepsilon \right) \right) \right) \\
        & \quad \times \left( 1 + \frac{r_{0}^{2}}{r^{2}} \left( 3 + \varepsilon (3 - 4 \cos^{2} \theta) \right) + \frac{r_{0}^{4}}{r^{4}} \left( 4 \varepsilon (1-\varepsilon) + 3 (1-\varepsilon)^{2} \right) + \frac{r_{0}^{6}}{r^{6}} (1-\varepsilon)^{2} (1+\varepsilon)
        \right)
        \Bigg]^{1/4}  \: .
    \end{split}
\end{equation}
Of course, in the limit $\varepsilon \rightarrow 0$, we recover the metric function for pure $AdS^{5} \times S^{5}$. One can demonstrate directly from the above expressions that these metric functions can be made uniformly close to those of pure $AdS^{5} \times S^{5}$
in $r \in [r_*, \infty)$ and $\theta \in [0, \frac{\pi}{2}]$ for sufficiently small $\varepsilon$; we have plotted some examples in Section \ref{sec:oneparam}.




\section{Space of solutions for the single pole case} \label{app:lksol}

In this section, we will understand the space of solutions to the constraints (\ref{eq:LKSingleStack}).

First, taking a linear combination of the last two equations, one obtains
\begin{equation} \label{eq:diophantine}
    N_{D5} L + N_{NS5} K = N + N_{D5} N_{NS5} \: ,
\end{equation}
so it is necessary that 
\begin{equation}
\label{GCD}
     G \equiv \textnormal{gcd}(N_{D5}, N_{NS5}) \mid N\: ,
\end{equation}

Choosing any $N_{NS5}$ and $N_{D5}$ satisfying this constraint, the linear diophantine equation (\ref{eq:diophantine}) for $K$ and $L$ will always have multiple integer solutions of the form
\begin{equation}
\label{mline}
    L = L_{0} + m \frac{N_{NS5}}{\textnormal{gcd}(N_{D5}, N_{NS5})} \: , \quad K = K_{0} - m \frac{N_{D5}}{\textnormal{gcd}(N_{D5}, N_{NS5})} \: , \quad m \in \mathbb{Z} \: ,
\end{equation}
with $(L_{0}, K_{0})$ some nominal solution.

There will be at least one solution for positive $K$ and $L$, since for real $m$, (\ref{mline}) parameterizes a line that intersects the positive quadrant of the $(K,L)$ plane, and the equal spacing between the $(K,L)$ values for integer $m$ is less than the length of the line segment in the positive quadrant:
\begin{equation}
     \sqrt{N_{D5}^{2} + N_{NS5}^{2}} < \sqrt{ \left( N_{D5} + \frac{N}{N_{NS5}} \right)^{2} + \left( N_{NS5} + \frac{N}{N_{D5}}  \right)^{2}}  \: .
\end{equation}
The number of solutions for $(K,L)$ is evidently of order
\be
\sqrt{ \left( N_{D5} + \frac{N}{N_{NS5}} \right)^{2} + \left( N_{NS5} + \frac{N}{N_{D5}}  \right)^{2} \over N_{D5}^{2} + N_{NS5}^{2}} \: 
\ee
so for $N_{D5},N_{NS5} \gg N$ we typically have only a single solution. The conditions that $K$ and $L$ are positive combined with (\ref{eq:diophantine}) mean that any solution will satisfy
\be
K < {N \over N_{NS5}} + N_{D5} \qquad \quad L < {N \over N_{D5}} + N_{NS5} \; .
\ee



Now, given any choice of $(N_{D5}, N_{NS5})$ satisfying (\ref{GCD}) and positive $(K,L)$ satisfying (\ref{eq:diophantine}), we will show that there is a unique positive $(k,l)$ satisfying the constraints (\ref{eq:LKSingleStack}). We do so by combining these constraints to yield
\be
{{K \over \nd} + {L \over \nn} - 1 \over {\nn \over g \nd} {k \over l} + 1} - {L \over \nn} + 1 = {2 \over \pi} \arctan{k \over l} \: .
\ee
The right side increases monotonically from 0 to 1 as $k/l$ increases from $0$ to $\infty$. The left side varies monotonically from $K/ \nd > 0$ at $k/l = 0$ to $1-L / \nn$ for large $k/l$. Thus, there is exactly one solution for $k/l$. Call this $k/l = m$.

We then have a unique solution $(k,l)$ that is the intersection between the line $k = ml$ and the line 
\be
{k \over \sqrt{g}} \nn + l \sqrt{g} \nd = N \; .
\ee

In terms of $m$, the result is
\beas
k &=& {N \over {N_{NS5} \over \sqrt{g}} + {\sqrt{g} N_{D5} \over m}} \cr
l &=& {N \over {m N_{NS5} \over \sqrt{g}} + \sqrt{g} N_{D5}} \cr \: .
\eeas

\section{General families with single D5-pole/NS5-pole and arbitrarily large \texorpdfstring{$AdS^{5} \times S^{5}$}{} region}  \label{sec:generalfam}

We will here provide a significant generalization to the one-parameter family initially introduced in Section \ref{sec:oneparam}. 
Our construction of a one-parameter family analogous to the one appearing in that section occurs most simply when $g$ is such that there exists $m \in \mathbb{N}^{+}$ with
\begin{equation} \label{eq:quantg}
    \arctan(m / g) = \frac{\pi}{2} \frac{a}{b} \: , \qquad a, b \in \mathbb{N}^{+} \: , \quad \textnormal{gcd}(a, b) = 1 \: , \quad \frac{a}{b} \in (0, 1) \: .
\end{equation}
That is, we have $g = \frac{m}{\tan \left( \frac{\pi}{2} \frac{a}{b} \right)}$, with $m, a, b$ positive integers and $0 < \frac{a}{b} < 1$ in reduced form. 
In this case, we will take
\begin{equation}
    \begin{split}
        N_{D5}(n) & = b f_{n} + \alpha \: , \qquad
        N_{NS5}(n) = b m f_{n} + \beta \: ,  \\
        L(n) & = a m f_{n} + \gamma \: , \qquad
        K(n) = (b - a) f_{n} + \delta \: ,  
    \end{split}
\end{equation}
where $f_{n}$ is a sequence which we leave undetermined for now. We then see that
\begin{equation} \label{eq:integersolvable}
    \begin{split}
        N_{D5}(n) L(n) + N_{NS5}(n) K(n) & = \left( b f_{n} + \alpha \right) \left( a m f_{n} + \gamma \right) + \left( b m f_{n} + \beta \right) \left( (b - a) f_{n} + \delta \right) \\
        & = N_{D5}(n) N_{NS5}(n) + \left( (a-b) m \alpha + b \gamma - a \beta + b m \delta \right) f_{n} \\
        & \qquad + \alpha \gamma + \beta \delta - \alpha \beta
    \end{split}
\end{equation}
so to ensure that (\ref{eq:diophantine}) holds, we would like to ask whether or not it is possible to choose $\alpha, \beta, \gamma, \delta$ such that
\begin{equation}
    \begin{split}
        0 & = (a-b) m \alpha + b \gamma - a \beta + b m \delta \\
        N & = \alpha \gamma + \beta \delta - \alpha \beta \: .
    \end{split}
\end{equation}
In fact, these equations \textit{are} solvable for any $(a, b, m)$. In particular, substituting the former into the latter yields
\begin{equation}
    \begin{split}
        N & = \left( \frac{(b-a)}{b} \alpha - \delta \right) \left( m \alpha - \beta \right)  \: .
    \end{split}
\end{equation}
If we take
\begin{equation}
    \beta = m \alpha - b \: ,
\end{equation}
then this equation gives
\begin{equation}
    (a-b) \alpha + b \delta = - N \: .
\end{equation}
We know that $\textnormal{gcd}\left( (a-b), b \right) = 1$, since $a$ and $b$ were chosen to be relatively prime, so this linear diophantine equation has an integer solution $(\alpha, \delta)$. We may then define 
\begin{equation}
    \gamma \equiv \frac{a}{b} \beta - \frac{a-b}{b} m \alpha - m \delta = -a + m (\alpha - \delta) \: ,
\end{equation}
which is manifestly integral.

We thus define the sequence of parameters $\left( N_{D5}(n), N_{NS5}(n), L(n), K(n) \right)$ by this choice $(\alpha, \beta, \gamma, \delta)$, taking $f_{n}$ to be any growing sequence. 
Since $\sqrt{g} N_{D5} l + \frac{1}{\sqrt{g}} N_{NS5} k = N$ implies that both $l$ and $k$ are at most $O\left( f_{n}^{-1} \right)$, the equations (\ref{eq:LKSingleStack}) yield
\begin{equation}
    \begin{split}
        \frac{L(n)}{N_{NS5}(n)} & = \frac{a}{b} + O \left( f_{n}^{-1} \right) = \frac{2}{\pi} \arctan(m/g) + O \left( f_{n}^{-1} \right) \\
        & = \frac{2}{\pi} \arctan(l/k) + O \left( f_{n}^{-2} \right) \: ,
    \end{split}
\end{equation}
and thus 
\begin{equation}
    l/k = m/g + O \left( f_{n}^{-1} \right) \: .
\end{equation}
It follows that
\begin{equation}
        |cl - dk| = \big| \left( \sqrt{g} b f_{n} \right) \left( \frac{km}{g} + O\left( f_{n}^{-2} \right) \right)  - \left( \frac{1}{\sqrt{g}} b m f_{n} \right) k \big| = O \left( f_{n}^{-1} \right) \: ,
\end{equation}
as desired. 

Thus, in the case that the string coupling $g$ satisfies (\ref{eq:quantg}), we are able to identify a one-parameter family with scaling
\begin{equation}
    m N_{D5} \sim N_{NS5} \sim \frac{b}{a} L \sim \frac{m b}{(b-a)} K \: .
\end{equation}
It is notable that such $g$ are dense in $\mathbb{R}^{+}$, since the map $\tan \frac{\pi}{2} \left( \cdot \right) : (0, 1) \rightarrow (0, \infty)$ is a continuous bijection, implying that the image of a dense set in this function is dense. We should therefore be able to extend the above result by considering sequences of suitable rational approximations.

Indeed, suppose that we fix arbitrary $g$ and take as ansatz the linear scaling
\begin{equation}
    z N_{D5} \sim N_{NS5} \: ,
\end{equation}
with $z \in \mathbb{R}^{+}$ \textit{any} fixed positive constant. 
In this case, requiring (\ref{eq:purecond}) to be satisfied implies
\begin{equation}
    \sqrt{g} l \sim \frac{z}{\sqrt{g}} k \: ,
\end{equation}
and given the relationship between linking numbers and SUGRA parameters (and the assumption that $l, k$ will be suppressed), this would appear to require
\begin{equation}
    \frac{L}{N_{NS5}} \sim \frac{2}{\pi} \arctan(z/g) \: , \qquad \frac{K}{N_{D5}} \sim \frac{2}{\pi} \arctan(g /z) \: .
\end{equation}
We would like to construct a sequence of parameters $\left( N_{D5}(n), N_{NS5}(n), L(n), K(n) \right)$ exhibiting the scaling that we have suggested, subject to the requirement that these parameters must be positive integers. The most natural way to approach this is to take sequences of rationals $\frac{a_{n}}{b_{n}}, \frac{p_{n}}{q_{n}}$ in reduced form such that
\begin{equation}
    \frac{a_{n}}{b_{n}} \rightarrow \frac{2}{\pi} \arctan(z / g) \: , \qquad \frac{p_{n}}{q_{n}} \rightarrow z \: ,
\end{equation}
and then define\footnote{In this section, $o()$ refers to the standard ``little $o$'' notation.} 
\begin{equation}
    \begin{split}
        & N_{D5}(n) = b_{n} q_{n} f_{n} + \alpha_{n} \: , \qquad \qquad \: \: \alpha_{n} = o(b_{n} q_{n} f_{n}) \\
        & N_{NS5}(n) = b_{n} p_{n} f_{n} + \beta_{n} \: , \qquad \qquad \beta_{n} = o(b_{n} p_{n} f_{n}) \\
        & L(n) = a_{n} p_{n} f_{n} + \gamma_{n} \: , \qquad \qquad \quad \: \: \gamma_{n} = o(a_{n} p_{n} f_{n}) \\
        & K(n) = (b_{n} - a_{n}) q_{n} f_{n} + \delta_{n} \: , \qquad \: \: \: \delta_{n} = o\left( (b_{n} - a_{n}) q_{n} f_{n}\right) \: ,
    \end{split}
\end{equation}
where $f_{n}$ is left undetermined for the time being. Equation (\ref{eq:diophantine}) then implies
\begin{equation}
    \begin{split}
        N_{D5}(n) L(n) + N_{NS5}(n) K(n) & = \left( b_{n} q_{n} f_{n} + \alpha_{n} \right) \left( a_{n} p_{n} f_{n} + \gamma_{n} \right) \\
        & \qquad + \left( b_{n} p_{n} f_{n} + \beta_{n} \right) \left( (b_{n} - a_{n}) q_{n} f_{n} + \delta_{n} \right) \\
        & = N_{D5}(n) N_{NS5}(n) \\
        & \qquad + \left( (a_{n}-b_{n}) p_{n} \alpha_{n} + b_{n} q_{n} \gamma_{n} - a_{n} q_{n} \beta_{n} + b_{n} p_{n} \delta_{n} \right) f_{n} \\
        & \qquad + \alpha_{n} \gamma_{n} + \beta_{n} \delta_{n} - \alpha_{n} \beta_{n} \: .
    \end{split}
\end{equation}
For any fixed $n$, this is precisely the same as (\ref{eq:integersolvable}), which we found to be consistent with the requirement $N_{D5} L + N_{NS5} K = N + N_{D5} N_{NS5}$ for suitably chosen $(\alpha, \beta, \gamma, \delta)$. Consequently, we may here find $(\alpha_{n}, \beta_{n}, \gamma_{n}, \delta_{n})$ which make our definitions of the parameters consistent with this equation for each $n$; once we have defined $(a_{n}, b_{n}, p_{n}, q_{n})$ and $(\alpha_{n}, \beta_{n}, \gamma_{n}, \delta_{n})$ in this way, we may then simply choose a sequence $f_{n}$ which scales sufficiently quickly such that we recover the necessary asymptotics
\begin{equation}
    \alpha_{n} = o(b_{n} q_{n} f_{n}) \: , \quad \beta_{n} = o(b_{n} p_{n} f_{n}) \: , \quad \gamma_{n} = o(a_{n} p_{n} f_{n}) \: , \quad \delta_{n} = o\left( (b_{n} - a_{n}) q_{n} f_{n}\right) \: .
\end{equation}
The sequence of solutions that we have defined will then have the desired asymptotic suppression of $\max\{l, k\}$ and $|cl - dk|$, as can be shown in a manner identical to that discussed above. 

\section{Nearby solutions with multiple poles}
\label{app:pert}

It is reasonable to expect that the precise form of our boundary condition, and in particular the linear quiver from which our boundary condition descends, can be relaxed somewhat, and indeed we expect that the broad geometrical features of the holographic description, including the existence of a large $AdS^{5} \times S^{5}$ region, should be robust to certain ``small" deformations of this quiver. As a concrete example, we may consider a family of solutions (the simplest family constructed earlier in Appendix \ref{sec:generalfam}) with parameters of the form
\begin{equation}
    N_{D5} = b n + \alpha \: , \quad N_{NS5} = b z n + \beta \: , \quad L = azn + \gamma \: , \quad K = (b-a) n + \delta \:  ;
\end{equation}
here, $\alpha, \beta, \gamma, \delta$ are constants chosen to satisfy $N = \alpha \gamma + \beta \delta - \alpha \beta$, and the constants $a, b, z$ satisfy
\begin{equation}
    \tan^{-1}(z/g) = \frac{\pi}{2} \frac{a}{b} \: , \quad a, b \in \mathbb{N}^{+} \: , \quad \textnormal{gcd}(a, b) = 1 \: .
\end{equation}
Each element of this sequence corresponds to a quiver of the form provided in Figure \ref{fig:LKQuiver}.
We will now consider deforming these quivers for each $n$ by coupling an additional $s(n)$ fundamental hypermultiplets to the $(L+1)^{\textnormal{th}}$ node of the quiver, where $s(n)$ may scale with $n$ but we require $s(n) = o(n)$. 
In this deformation, we have two stacks of D5-branes and two stacks of NS5-branes with inequivalent linking numbers, described by the parameters
\begin{equation}
        N_{D5}^{(1)} = b n + \alpha \: , \quad N_{D5}^{(2)} = s \: , \quad N_{NS5}^{(1)} = a z n + \gamma + 1 \: , \quad N_{NS5}^{(2)} = (b - a) z n + \beta - \gamma - 1 \: ,
\end{equation}
and
\begin{equation}
    L_{1} = a z n + \gamma \: , \quad L_{2} = a z n + \gamma + 1 \: , \quad K_{1} = (b - a) n + \delta \: , \quad K_{2} = (b - a) n + \delta + s \: .
\end{equation}
At leading order (namely at order $O(n)$), (\ref{KLkl}) gives the conditions
\begin{equation}
    \begin{split}
        \frac{\pi}{2} a & = a \tan^{-1}(l_{1} / k_{1}) + (b-a) \tan^{-1}(l_{1} / k_{2})  + o(n^{0}) \\
        \frac{\pi}{2} a & = a \tan^{-1}(l_{2} / k_{1}) + (b-a) \tan^{-1}(l_{2} / k_{2}) + o(n^{0}) \\
        \frac{\pi}{2} (b-a) & = b \tan^{-1}(k_{1} / l_{1})  + o(n^{0}) \\
        \frac{\pi}{2} ( b - a) & = b \tan^{-1}(k_{2} / l_{1})  + o(n^{0}) \: ,
    \end{split}
\end{equation}
from which we can infer
\begin{equation}
    \frac{g}{z} = \frac{k_{1}}{l_{1}} + o(n^{0}) = \frac{k_{2}}{l_{1}} + o(n^{0}) = \frac{k_{1}}{l_{2}} + o(n^{0}) = \frac{k_{2}}{l_{2}} + o(n^{0}) 
\end{equation}
and thus from (\ref{Nsum})
\begin{equation}
\begin{split}
    l_{1} & = \frac{N}{\sqrt{g} b} \frac{1}{2n} + o(n^{-1}) \: , \qquad \: l_{2} = \frac{N}{\sqrt{g} b} \frac{1}{2n} + o(n^{-1}) \: , \\ k_{1} & = \frac{\sqrt{g} N}{z b} \frac{1}{2n} + o(n^{-1}) \: , \qquad \: k_{2} = \frac{\sqrt{g} N}{z b} \frac{1}{2n}+ o(n^{-1}) \: ,
\end{split}
\end{equation}
and
\begin{equation}
    \Delta = \big| c_{1} l_{1} + c_{2} l_{2} - d_{1} k_{1} - d_{2} k_{2} \big| = o(n^{0}) \: .
\end{equation}
Since $\max\{l_{A}, k_{B}\}$ and $\Delta$ are again suppressed for large $n$, we find that we recover the desired geometrical features in this limit. In particular, while we now have two D5-brane throats and two NS5-brane throats, the total D5-brane and NS5-brane charges are approximately the same as before, and the separation between each pair of 5-brane throats in this case is subleading in $n$,
\begin{equation}
    \frac{l_{1} - l_{2}}{l_{1}} = o(n^{0}) \: , \qquad \frac{k_{1} - k_{2}}{k_{1}} = o(n^{0}) \: .
\end{equation}

It is straightforward to show that a similar argument can be applied to a more general version of this deformation, where we couple $o(n)$ fundamental hypermultiplets at each of $O(n^{0})$ nodes in the quiver, where the location of these nodes relative to the left endpoint of the quiver scales proportionally to the overall size of the quiver with $n$. 

Another deformation of interest involves coupling an additional small quiver to the left endpoint of our initial quiver, i.e. the endpoint opposite that which is coupled directly to the 4D theory upon imposing our field theory boundary condition. Here, ``small quiver" refers to a quiver described by an $O(1)$ number of parameters $(N_{D5}^{(A)}, L_{A})$ and $(N_{NS5}^{(B)}, K_{B})$, all of which are dominated by our initial parameters $(N_{D5}, N_{NS5}, L, K)$. We can couple the large and small quivers together via bifundamental matter coupled to an extra $U(m)$ node, where $m$ is also dominated by our initial parameters; the result will be a good quiver, provided that the small quiver is good. This procedure results in a boundary condition described by many distinct parameters, which we can denote by $(N_{D5}^{(A)}, L_{A})_{A=1 \ldots p}$ and $(N_{NS5}^{(B)}, K_{B}))_{B=1 \ldots q}$ with some abuse of notation (they are different from those describing the small quiver). 
Notably, $(N_{D5}^{(p)}, N_{NS5}^{(q)}, L_{p}, K_{q})$ agree with the original parameters $(N_{D5}, N_{NS5}, L, K)$ at leading order. From
\begin{equation}
\begin{split}
    L_{p} & = \sqrt{g} l_{p} + \frac{2}{\pi} \sum_{B} N_{NS5}^{(B)} \arctan \left( l_{p} / k_{B} \right) \\
     K_{q} & = \frac{1}{\sqrt{g}} k_{q} + \frac{2}{\pi} \sum_{A} N_{D5}^{(A)} \arctan \left( k_{q} / l_{A} \right) \: ,
\end{split}    
\end{equation}
and the fact that $N_{D5}^{(A)}, N_{NS5}^{(B)} \ll N_{D5}^{(p)}, N_{NS5}^{(q)}$ for $A < p$ and $B < q$, we see that the leading behaviour of $l_{p}, k_{q}$ will be the same as before the deformation. Moreover, the remaining equations for the linking numbers imply
\begin{equation}
    l_{A} / k_{q} = O \left( L_{A} / N_{NS5}^{(q)} \right) \: , \qquad k_{B} / l_{p} = O \left( K_{B} / N_{D5}^{(p)} \right) 
\end{equation}
for $A < p$ and $B < q$. Consequently, the newly added parameters are suppressed compared to $l, k$, and contribute to $\Delta$ at subleading order; we therefore arrive again at a solution with a large $AdS^{5} \times S^{5}$ region.


\section{Multi-wedge generalizations} \label{app:multi}

\begin{figure}
    \centering
    \includegraphics[height=8cm]{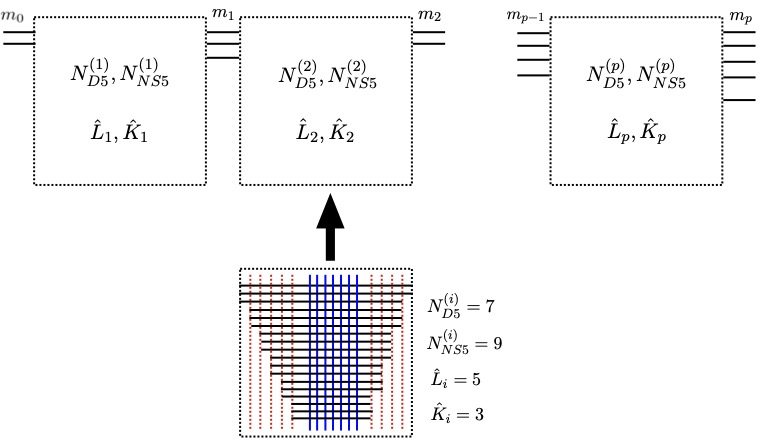}
    \caption{
    D-brane construction giving rise to the class of boundary conditions considered in this appendix. We have ``blocks" consisting of D3-branes stretched between $N_{D5}^{(i)}$ D5-branes and $N_{NS5}^{(i)}$ NS5-branes with fixed linking numbers $L_{i}, K_{i}$, where ultimately we will take $N_{D5}^{(i)}, N_{NS5}^{(i)}, L_{i}, K_{i}$ to scale with some large quantity. 
    The $(i-1)^{\textnormal{th}}$ and $i^{\textnormal{th}}$ blocks are connected by $m_{i}$ D3-branes. We give an example of the brane configuration in one such block, with D3-branes shown in black, D5-branes in blue, and NS5-branes in red. }
    \label{fig:manybrane}
\end{figure}

Our goal in this section is to understand how to construct theories whose holographic description involves several wedges of $AdS^{5} \times S^{5}$ connected by interface branes; this applies to the BCFT case as well as the case involving 3D SCFTs which descend from linear or circular quiver gauge theories. The intuition behind our construction is illustrated in Figure \ref{fig:manybrane}.

Our construction in this section will begin with a list
\begin{equation}
    (m_{0}, m_{1}, m_{2}, \ldots, m_{p-1}, m_{p})
\end{equation}
of non-negative integers, where we fix $p$ for concreteness. In the linear quiver case, we will have $m_{0} = m_{p} = 0$, in the circular quiver case, we will have $m_{0} = m_{0} = L \neq 0$, and in the BCFT case, we will have $m_{0} = 0$ and $m_{p} = N$. 
We would then like to define the required field theory data
\begin{equation}
    (L_{1}, \ldots, L_{p}) \: , \: (K_{1}, \ldots, K_{p}) \: , \: (N_{D5}^{(1)}, \ldots, N_{D5}^{(p)}) \: , \: (N_{NS5}^{(1)}, \ldots, N_{NS5}^{(p)}) \: ,
\end{equation}
where the linking numbers are listed in increasing order. 
We will define these via the brane configuration depicted in Figure \ref{fig:manybrane}; we have ``blocks" with large numbers of D5-branes and NS5-branes $N_{D5}^{(A)},  N_{NS5}^{(A)}$, each with large linking numbers $L_{A}, K_{A}$ respectively, and the $(A-1)^{\textnormal{th}}$ and $A^{\textnormal{th}}$ blocks are connected by $m_{A}$ D3-branes.
The quantities $(N_{D5}^{(A)}, N_{NS5}^{(A)}, L_{A}, K_{A})$ which parametrize the $A^{\textnormal{th}}$ block may be constructed in a completely identical manner to the construction of the one-parameter families we considered in Section \ref{sec:oneparam} and Appendix \ref{sec:generalfam}, with the simple replacement $N \rightarrow \left( m_{A} - m_{A-1} \right)$; in particular, the linking numbers $\bar{L}_{A}, \bar{K}_{A}$ that we would obtain from that construction will be related to the correct linking numbers $L_{A}, K_{A}$ in the full quiver of the present construction by
\begin{equation} \label{eq:fulllinking}
    L_{A} = \bar{L}_{A} + \sum_{B = 1}^{A-1} N_{NS5}^{(B)} \: , \qquad K_{A} = \bar{K}_{A} + \sum_{B = 1}^{A-1} N_{D5}^{(B)} \: ,
\end{equation}
since we need to account for the fact that the linking numbers depend on the quantities of 5-branes present in previous blocks. 
Ultimately, we will take all of the $m_{A}$ (and the number of blocks $p$) to be $O(1)$ in some large parameters 
which will determine the number of 5-branes and linking numbers in the $A^{\textnormal{th}}$ block.

The above is the sense in which these boundary conditions correspond to ``glued together" sub-quivers; the sub-quivers that are being coupled in this case are precisely those that arose in the discussion of Section \ref{sec:oneparam}, corresponding to boundary conditions described by single linking numbers $L, K$, with the replacement $N \rightarrow \left( m_{A} - m_{A-1} \right)$ in the present context.

We proceed to define $(\bar{L}_{A}, \bar{K}_{A}, N_{D5}^{(A)}, N_{NS5}^{(A)})$, beginning in full generality with the case of arbitrary coupling $g$; in general, we construct these exactly as in 
Appendix \ref{sec:generalfam}, taking
\begin{equation}
    N_{D5}^{(A)} = b_{n}^{(A)} q_{n}^{(A)} f_{n}^{(A)} + \alpha_{n}^{(A)} \: , \qquad N_{NS5}^{(A)} = b_{n}^{(A)} p_{n}^{(A)} f_{n}^{(A)} + \beta_{n}^{(A)} 
\end{equation}
and
\begin{equation}         \bar{L}_{A} = a_{n}^{(A)} p_{n}^{(A)} f_{n}^{(A)} + \gamma_{n}^{(A)} \: , \qquad \bar{K}_{A} = (b_{n}^{(A)}-a_{n}^{(A)}) q_{n}^{(A)} f_{n}^{(A)} + \delta_{n}^{(A)}
\end{equation}
where 
\begin{equation}
    \frac{a_{n}^{(A)}}{b_{n}^{(A)}} \rightarrow \frac{2}{\pi} \tan^{-1}(z_{A} / g) \: , \qquad \frac{p_{n}^{(A)}}{q_{n}^{(A)}} \rightarrow z_{A}
\end{equation}
for some $z_{A}$, the quantities
$\alpha_{n}^{(A)}, \beta_{n}^{(A)}, \gamma_{n}^{(A)}, \delta_{n}^{(A)}$ sastisfy
\begin{equation}
    \alpha_{n}^{(A)} \gamma_{n}^{(A)} + \beta_{n}^{(A)} \delta_{n}^{(A)} - \alpha_{n}^{(A)} \beta_{n}^{(A)} = m_{A} - m_{A-1} \: ,
\end{equation} 
and $f_{n}^{(A)}$ is quickly-scaling.
Then, passing to the linking numbers by (\ref{eq:fulllinking}),
we have that
\begin{equation}
    \sum_{A = 1}^{p} \left( N_{D5}^{(A)} L_{A} + N_{NS5}^{(A)} K_{A} \right) = m_{p} - m_{0} + N_{D5} N_{NS5}
\end{equation}
and the linking numbers are increasing
by construction, 
We will also require that $f_{n}^{(A)}$ scales sufficiently quickly relative to $f_{n}^{(A-1)}$ such that the parameters in block $A$ scale at least as quickly as the parameters in block $A-1$. 
We can now consider how the SUGRA parameters behave for each case. 

\subsection{Multi-wedge dual of BCFT}

Recalling that
\begin{equation}
    \sum_{A=1}^{p} \left( \sqrt{g} N_{D5}^{(A)} l_{A} + \frac{1}{\sqrt{g}} N_{NS5}^{(A)} k_{A} \right) = N \: ,
\end{equation}
and all of the $l_{A}, k_{A}$ are positive, we see that one must have
\begin{equation}
    l_{A} < \frac{N}{N_{D5}^{(A)}} \: , \qquad k_{A} < \frac{N}{N_{NS5}^{(A)}} \: ,
\end{equation}
so that in particular
\begin{equation}
    \lim_{n \rightarrow \infty} l_{A}, k_{A} = 0 \: .
\end{equation}
We have from our definitions
\begin{equation}
    \begin{split}
        L_{A} & = a_{n}^{(A)} p_{n}^{(A)} f_{n}^{(A)} + \gamma_{n}^{(A)} + \sum_{B=1}^{A-1} \left( b_{n}^{(A)}  p_{n}^{(A)} f_{n}^{(A)} + \beta_{n}^{(A)} \right) \\
        K_{A} & = (b_{n}^{(A)} - a_{n}^{(A)}) q_{n}^{(A)} f_{n}^{(A)} + \delta_{n}^{(A)} + \sum_{B=1}^{A-1} \left( b_{n}^{(A)}  q_{n}^{(A)} f_{n}^{(A)} + \alpha_{n}^{(A)} \right) \: ,
    \end{split}
\end{equation}
as well as the relations to SUGRA parameters
\begin{equation}
    \begin{split}
        L_{A} & = \sqrt{g} l_{A} + \frac{2}{\pi} \sum_{B=1}^{p} \left( b_{n}^{(A)}  p_{n}^{(A)} f_{n}^{(A)} + \beta_{n}^{(A)} \right) \tan^{-1} \left( l_{A} / k_{B} \right) \\
        K_{A} & = \frac{1}{\sqrt{g}} k_{A} + \frac{2}{\pi} \sum_{B=1}^{p} \left( b_{n}^{(A)}  q_{n}^{(A)} f_{n}^{(A)} + \alpha_{n}^{(A)} \right)  \tan^{-1} \left( k_{A} / l_{B} \right) \: .
    \end{split}
\end{equation}

Comparing these expressions at leading order, we see that consistency is achieved by requiring
\begin{equation}
    \lim_{n \rightarrow \infty} \frac{l_{A}}{k_{A}}
    = \frac{z_{A}}{g} 
\end{equation}
and
\begin{equation}
    \lim_{n \rightarrow \infty} \frac{l_{A}}{l_{B}} = \lim_{n \rightarrow \infty} \frac{k_{A}}{k_{B}} = 0 \: , \qquad A < B \: .
\end{equation}
Schematically, we can say that $g l_{A} \sim z_{A} k_{A}$
and
\begin{equation}
    l_{1} \ll l_{2} \ll \ldots \ll l_{p} \ll 1 \: , \quad k_{1} \ll k_{2} \ll \ldots \ll k_{p} \ll 1 \: .
\end{equation}
We therefore find
\begin{equation}
    \lim_{n \rightarrow \infty} \big| \sum_{A=1}^{p} \left( N_{D5}^{(A)} l_{A} - N_{NS5}^{(A)} k_{A} \right) \big| =  0  \: ,
\end{equation}
as desired.

We have demonstrated that our construction thus far possesses a large $AdS^{5} \times S^{5}$ asymptotic region; to ensure that we recover a multi-wedge deep in the interior, we will actually consider a subset of the families defined so far for which the parameters $(N_{D5}^{(A)}, N_{NS5}^{(A)}, \bar{L}_{A}, \bar{K}_{A})$ of block $A$ are all taken to scale with the same large parameter as the parameters of block $A-1$, as opposed to scaling strictly faster. Note that the ``doubled" construction of Section \ref{sec:3dwedge} is an example of this choice. 
In this case, it suffices to note that for $l_{A}, k_{A} \ll r \ll l_{A+1}, k_{A+1}$, we find the leading behaviour of $h_{1}, h_{2}$ to be
\bea
h_1 &=&{\pi \ell_s^2 \over 2} {r \cos \theta \over \sqrt{g}} + {\ell_s^2 \over 4}\sum_A {c_A \over  \sqrt{g}} \ln \left( {(r \cos \theta + l_A)^2 +  r^2 \sin^2 \theta \over (r \cos \theta-l_A)^2 + r^2 \sin^2 \theta} \right) \cr
&=& \ell_s^2 \sum_{B \leq A} N_{D5}^{(B)} \left( \frac{l_{B}}{r} \cos \theta + O(l_{B}^{3} / r^{3}) \right) + \ell_s^2 \sum_{B > A} N_{D5}^{(B)} \left( \frac{r}{l_{B}} \cos \theta + O(r^{3} / l_{B}^{3} ) \right) \cr
&\approx& \ell_s^2 \cos \theta \left( N_{D5}^{(A)}  \frac{l_{A}}{r} +  N_{D5}^{(A+1)} \frac{r}{l_{A+1}}  \right) 
\eea
and
\bea
\label{gensol}
h_2 &=& {\pi \ell_s^2 \over 2} \sqrt{g} r \sin \theta + {\ell_s^2 \over 4}\sum_A d_A \sqrt{g} \ln \left( {r^2 \cos^2 \theta + (r \sin \theta + k_A)^2 \over r^2 \cos^2 \theta + (r \sin \theta-k_A)^2} \right) \cr
&=& \ell_s^2 \sum_{B \leq A} N_{NS5}^{(B)} \left( \frac{k_{B}}{r} \sin \theta + O(k_{B}^{3} / r^{3}) \right) + \ell_s^2 \sum_{B > A} N_{NS5}^{(B)} \left( \frac{r}{k_{B}} \cos \theta + O(r^{3} / k_{B}^{3} ) \right) \cr
&\approx& \ell_s^2 \sin \theta \left( N_{NS5}^{(A)}  \frac{k_{A}}{r} +  N_{NS5}^{(A+1)} \frac{r}{k_{A+1}}  \right)  \; .
\eea
Since $z_{B} N_{D5}^{(B)} \sim N_{NS5}^{(B)}$ and $\sqrt{g} l_{B} \sim \frac{z_{B}}{\sqrt{g}} k_{B}$, the geometry in this region is approximately that of $AdS^{5} \times S^{5}$, where the value of $r_{0}$ is proportional to the geometric mean of $l_{A}$ (or $k_{A}$) and $l_{A+1}$ (or $k_{A+1}$), and the AdS radius in this wedge scales relative to the AdS radius in the asymptotic region as $L_{\textnormal{wedge}}^{4} / L^{4} \sim \frac{N_{D5}^{(A)} N_{D5}^{(A+1)}}{N^{2}} \frac{l_{A}}{l_{A+1}}$.

\subsection{Multi-wedge dual of SCFT: linear quiver}

As at the end of last section, we will continue to restrict to the case where the linking numbers and charges for each block are all taken to scale with the same large parameter.
The linking numbers $L_{A}, K_{A}$ are related to parameters $N_{3}^{(A)}, \hat{N}_{3}^{(A)}$ by
\begin{equation}
    N_{3}^{(A)} = N_{NS5} - L_{A} \: , \qquad  \hat{N}_{3}^{(A)} = K_{A} \: ,
\end{equation}
so we can write (\ref{eq:3dquantization}) as
\begin{equation}
    \begin{split}
        L_{A} & = \frac{2}{\pi} \sum_{B} \hat{N}_{5}^{(B)} \tan^{-1} \left( e^{\hat{\delta}_{B} - \delta_{A} } \right) \\
        K_{B} & = \frac{2}{\pi} \sum_{A} N_{5}^{(A)} \tan^{-1} \left( e^{ \delta_{A} - \hat{\delta}_{B}} \right) \: . 
    \end{split}
\end{equation}
It is immediate that we obtain the desired behaviour in this case, since this system of equations is identical to the system from the BCFT case up to sub-leading terms if we identify $l_{A} \leftrightarrow e^{- \delta_{A}}$ and $k_{B} \leftrightarrow e^{- \hat{\delta}_{B}}$, and the definitions of $h_{1}, h_{2}$ will have the same leading behaviour in the regions of interest. 

\subsection{Multi-wedge dual of SCFT: circular quiver}

The solutions of type IIB supergravity describing the vacuum states of 3D SCFTs arising from circular quiver gauge theories have not yet been discussed in this note, but were first analyzed in \cite{Assel:2012cj}. These solutions are similar to those arising from linear quivers, with harmonic functions $h_{1}, h_{2}$ now given by
\begin{equation}
    \begin{split}
        h_{1} & = - \sum_{a=1}^{p} \gamma_{a} \ln \left( \prod_{n = - \infty}^{\infty} \tanh \left( \frac{\pi i}{4} - \frac{z - (\delta_{a} + 2nt)}{2} \right) \right) + \textnormal{c.c.} \: , \\
        h_{2} & = - \sum_{b=1}^{\hat{p}} \hat{\gamma}_{b} \ln \left( \prod_{n = - \infty}^{\infty} \tanh \left( \frac{\pi i}{4} - \frac{z - (\hat{\delta}_{b} + 2nt)}{2} \right) \right) + \textnormal{c.c.} \: ,
    \end{split}
\end{equation}
where $t$ is a positive parameter satisfying $0 \leq \delta_{a}, \hat{\delta}_{b} \leq 2t$. These functions are periodic under $\textnormal{Re}(z) \rightarrow \textnormal{Re}(z) + 2 t$ by construction, and we can alternatively express them using Jacobi $\vartheta$-functions as
\begin{equation}
    \begin{split}
        h_{1} & = - \sum_{a=1}^{p} \gamma_{a} \ln \left( \frac{\vartheta_{1}(\nu_{a} | \tau)}{\vartheta_{2}(\nu_{a}|\tau)} \right) + \textnormal{c.c.} \: , \qquad i \nu_{a} = - \frac{z - \delta_{a}}{2 \pi} + \frac{i}{4} \: , \\
        h_{2} & = - \sum_{b=1}^{\hat{p}} \hat{\gamma}_{b} \ln \left( \frac{\vartheta_{1}(\hat{\nu}_{b} | \tau)}{\vartheta_{2}(\hat{\nu}_{b}|\tau)} \right) + \textnormal{c.c.} \: , \qquad i \hat{\nu}_{b} = \frac{z - \hat{\delta}_{b}}{2 \pi} 
    \end{split}
\end{equation}
on a torus with modular parameter $\tau = i t / \pi$. 

The linking numbers and supergravity parameters are now related by
\begin{equation}
    \begin{split}
        L_{A} & = \frac{2}{\pi} \sum_{B} N_{NS5}^{(B)} \left( \sum_{n = 0}^{\infty} \arctan \left( e^{- \hat{\delta}_{B} + \delta_{A} - 2 n t} \right) - \sum_{n = 1}^{\infty} \arctan \left( e^{\hat{\delta}_{B} - \delta_{A} - 2 n t}  \right) \right) \\
        K_{A} & = \frac{2}{\pi} \sum_{B} N_{D5}^{(B)} \left( \sum_{n = 0}^{\infty} \arctan \left( e^{\hat{\delta}_{B} - \delta_{A} - 2 n t}  \right) - \sum_{n = 1}^{\infty} \arctan \left( e^{- \hat{\delta}_{B} + \delta_{A} - 2 n t} \right) \right)
    \end{split}
\end{equation}
and
\begin{equation}
    L = \frac{2}{\pi} \sum_{A} \sum_{B} N_{D5}^{(A)} N_{NS5}^{(B)} \sum_{s=1}^{\infty} s \left( \arctan(e^{\hat{\delta}_{B} - \delta_{A} - 2 s t}) + \arctan(e^{\delta_{A} - \hat{\delta}_{B} - 2 s t}) \right) \: ,
\end{equation}
where $L = m_{0} = m_{p}$. 

The linking number conditions can again be satisfied by requiring
\begin{equation}
    e^{-\delta_{A}}, e^{-\hat{\delta}_{A}} \ll e^{- \delta_{A+1}}, e^{- \hat{\delta}_{A+1}}
\end{equation} 
and
\begin{equation}
    e^{ \hat{\delta}_{A} - \delta_{A}} \sim  \tan \left( \frac{\pi}{2} \frac{\bar{K}_{A}}{N_{D5}^{(A)}} \right) \sim \frac{g}{z_{A}} \: ,
\end{equation}
provided $t$ is sufficiently large that
\begin{equation}
    e^{- \hat{\delta}_{B} + \delta_{A} - 2 t} \ll 1 \: , \qquad e^{\hat{\delta}_{B} - \delta_{A} - 2t} \ll 1 
\end{equation}
for all $A, B$. 
It is clear from the expression for $L$ that these conditions must be true, since $N_{D5}^{(A)}, N_{NS5}^{(B)} \gg L$. 
Again, in the region $\delta_{A} \ll \textnormal{Re}(z) \ll \delta_{A+1}$, the harmonic functions $h_{1}, h_{2}$ agree with those from the linear quiver case at leading order, since the additional contributions coming from the $n \neq 0$ terms will be suppressed.


\begin{thebibliography}{1}

\bibitem{Maartens:2010ar}
R.~Maartens and K.~Koyama, \emph{{Brane-World Gravity}},
  \href{http://dx.doi.org/10.12942/lrr-2010-5}{\emph{Living Rev. Rel.} {\bf 13}
  (2010) 5}, [\href{https://arxiv.org/abs/1004.3962}{{\tt 1004.3962}}].

\bibitem{Cooper:2018cmb}
S.~Cooper, M.~Rozali, B.~Swingle, M.~Van~Raamsdonk, C.~Waddell and D.~Wakeham,
  \emph{{Black Hole Microstate Cosmology}},
  \href{http://dx.doi.org/10.1007/JHEP07(2019)065}{\emph{JHEP} {\bf 07} (2019)
  065}, [\href{https://arxiv.org/abs/1810.10601}{{\tt 1810.10601}}].

\bibitem{Antonini:2019qkt}
S.~Antonini and B.~Swingle, \emph{{Cosmology at the end of the world}},
  \href{http://dx.doi.org/10.1038/s41567-020-0909-6}{\emph{Nature Phys.} {\bf
  16} (2020) 881--886}, [\href{https://arxiv.org/abs/1907.06667}{{\tt
  1907.06667}}].

\bibitem{VanRaamsdonk:2020tlr}
M.~Van~Raamsdonk, \emph{{Comments on wormholes, ensembles, and cosmology}},
  \href{https://arxiv.org/abs/2008.02259}{{\tt 2008.02259}}.

\bibitem{VanRaamsdonk:2021qgv}
M.~Van~Raamsdonk, \emph{{Cosmology from confinement?}},
  \href{https://arxiv.org/abs/2102.05057}{{\tt 2102.05057}}.

\bibitem{Almheiri:2019hni}
A.~Almheiri, R.~Mahajan, J.~Maldacena and Y.~Zhao, \emph{{The Page curve of
  Hawking radiation from semiclassical geometry}},
  \href{http://dx.doi.org/10.1007/JHEP03(2020)149}{\emph{JHEP} {\bf 03} (2020)
  149}, [\href{https://arxiv.org/abs/1908.10996}{{\tt 1908.10996}}].

\bibitem{Rozali:2019day}
M.~Rozali, J.~Sully, M.~Van~Raamsdonk, C.~Waddell and D.~Wakeham,
  \emph{{Information radiation in BCFT models of black holes}},
  \href{http://dx.doi.org/10.1007/JHEP05(2020)004}{\emph{JHEP} {\bf 05} (2020)
  004}, [\href{https://arxiv.org/abs/1910.12836}{{\tt 1910.12836}}].

\bibitem{Chen:2020uac}
H.~Z. Chen, R.~C. Myers, D.~Neuenfeld, I.~A. Reyes and J.~Sandor,
  \emph{{Quantum Extremal Islands Made Easy, Part I: Entanglement on the
  Brane}}, \href{http://dx.doi.org/10.1007/JHEP10(2020)166}{\emph{JHEP} {\bf
  10} (2020) 166}, [\href{https://arxiv.org/abs/2006.04851}{{\tt 2006.04851}}].

\bibitem{Chen:2020hmv}
H.~Z. Chen, R.~C. Myers, D.~Neuenfeld, I.~A. Reyes and J.~Sandor,
  \emph{{Quantum Extremal Islands Made Easy, Part II: Black Holes on the
  Brane}}, \href{http://dx.doi.org/10.1007/JHEP12(2020)025}{\emph{JHEP} {\bf
  12} (2020) 025}, [\href{https://arxiv.org/abs/2010.00018}{{\tt 2010.00018}}].

 \bibitem{Geng:2020qvw}
H.~Geng and A.~Karch, \emph{{Massive islands}},
  \href{http://dx.doi.org/10.1007/JHEP09(2020)121}{\emph{JHEP} {\bf 09} (2020)
  121}, [\href{https://arxiv.org/abs/2006.02438}{{\tt 2006.02438}}].

\bibitem{Geng:2020fxl}
H.~Geng, A.~Karch, C.~Perez-Pardavila, S.~Raju, L.~Randall, M.~Riojas et~al.,
  \emph{{Information Transfer with a Gravitating Bath}},
  \href{https://arxiv.org/abs/2012.04671}{{\tt 2012.04671}}.

\bibitem{Uhlemann:2021nhu}
C.~F. Uhlemann, \emph{{Islands and Page curves in 4d from Type IIB}},
  \href{https://arxiv.org/abs/2105.00008}{{\tt 2105.00008}}.

\bibitem{Randall:1999vf}
L.~Randall and R.~Sundrum, \emph{{An Alternative to compactification}},
  \href{http://dx.doi.org/10.1103/PhysRevLett.83.4690}{\emph{Phys. Rev. Lett.}
  {\bf 83} (1999) 4690--4693},
  [\href{https://arxiv.org/abs/hep-th/9906064}{{\tt hep-th/9906064}}].

\bibitem{Karch:2000ct}
A.~Karch and L.~Randall, \emph{{Locally localized gravity}},
  \href{http://dx.doi.org/10.1088/1126-6708/2001/05/008}{\emph{JHEP} {\bf 05}
  (2001) 008}, [\href{https://arxiv.org/abs/hep-th/0011156}{{\tt
  hep-th/0011156}}].

\bibitem{Karch:2000gx}
A.~Karch and L.~Randall, \emph{{Open and closed string interpretation of SUSY
  CFT's on branes with boundaries}},
  \href{http://dx.doi.org/10.1088/1126-6708/2001/06/063}{\emph{JHEP} {\bf 06}
  (2001) 063}, [\href{https://arxiv.org/abs/hep-th/0105132}{{\tt
  hep-th/0105132}}].

\bibitem{Takayanagi:2011zk}
T.~Takayanagi, \emph{{Holographic Dual of BCFT}},
  \href{http://dx.doi.org/10.1103/PhysRevLett.107.101602}{\emph{Phys. Rev.
  Lett.} {\bf 107} (2011) 101602}, [\href{https://arxiv.org/abs/1105.5165}{{\tt
  1105.5165}}].

\bibitem{Aharony:2011yc}
O.~Aharony, L.~Berdichevsky, M.~Berkooz and I.~Shamir, \emph{{Near-horizon
  solutions for D3-branes ending on 5-branes}},
  \href{http://dx.doi.org/10.1103/PhysRevD.84.126003}{\emph{Phys. Rev.} {\bf
  D84} (2011) 126003}, [\href{https://arxiv.org/abs/1106.1870}{{\tt
  1106.1870}}].

\bibitem{Assel:2011xz}
B.~Assel, C.~Bachas, J.~Estes and J.~Gomis, \emph{{Holographic Duals of D=3 N=4
  Superconformal Field Theories}},
  \href{http://dx.doi.org/10.1007/JHEP08(2011)087}{\emph{JHEP} {\bf 08} (2011)
  087}, [\href{https://arxiv.org/abs/1106.4253}{{\tt 1106.4253}}].

\bibitem{Bachas:2018zmb}
C.~Bachas and I.~Lavdas, \emph{{Massive Anti-de Sitter Gravity from String
  Theory}}, \href{http://dx.doi.org/10.1007/JHEP11(2018)003}{\emph{JHEP} {\bf
  11} (2018) 003}, [\href{https://arxiv.org/abs/1807.00591}{{\tt 1807.00591}}].

\bibitem{ArkaniHamed:2001ca}
N.~Arkani-Hamed, A.~G. Cohen and H.~Georgi, \emph{{(De)constructing
  dimensions}},
  \href{http://dx.doi.org/10.1103/PhysRevLett.86.4757}{\emph{Phys. Rev. Lett.}
  {\bf 86} (2001) 4757--4761},
  [\href{https://arxiv.org/abs/hep-th/0104005}{{\tt hep-th/0104005}}].

\bibitem{Bachas:2017rch}
C.~Bachas and I.~Lavdas, \emph{{Quantum Gates to other Universes}},
  \href{http://dx.doi.org/10.1002/prop.201700096}{\emph{Fortsch. Phys.} {\bf
  66} (2018) 1700096}, [\href{https://arxiv.org/abs/1711.11372}{{\tt
  1711.11372}}].

\bibitem{Akal:2020wfl}
I.~Akal, Y.~Kusuki, T.~Takayanagi and Z.~Wei, \emph{{Codimension two holography
  for wedges}},
  \href{http://dx.doi.org/10.1103/PhysRevD.102.126007}{\emph{Phys. Rev. D} {\bf
  102} (2020) 126007}, [\href{https://arxiv.org/abs/2007.06800}{{\tt
  2007.06800}}].

\bibitem{Bachas:2011xa}
C.~Bachas and J.~Estes, \emph{{Spin-2 spectrum of defect theories}},
  \href{http://dx.doi.org/10.1007/JHEP06(2011)005}{\emph{JHEP} {\bf 06} (2011)
  005}, [\href{https://arxiv.org/abs/1103.2800}{{\tt 1103.2800}}].

\bibitem{VanRaamsdonk:2020djx}
M.~Van~Raamsdonk and C.~Waddell, \emph{{Holographic and localization
  calculations of boundary F for $ \mathcal{N} $ = 4 SUSY Yang-Mills theory}},
  \href{http://dx.doi.org/10.1007/JHEP02(2021)222}{\emph{JHEP} {\bf 02} (2021)
  222}, [\href{https://arxiv.org/abs/2010.14520}{{\tt 2010.14520}}].

\bibitem{Gaiotto:2008sa}
D.~Gaiotto and E.~Witten, \emph{{Supersymmetric Boundary Conditions in N=4
  Super Yang-Mills Theory}},
  \href{http://dx.doi.org/10.1007/s10955-009-9687-3}{\emph{J. Statist. Phys.}
  {\bf 135} (2009) 789--855}, [\href{https://arxiv.org/abs/0804.2902}{{\tt
  0804.2902}}].

\bibitem{Gaiotto:2008ak}
D.~Gaiotto and E.~Witten, \emph{{S-Duality of Boundary Conditions In N=4 Super
  Yang-Mills Theory}},
  \href{http://dx.doi.org/10.4310/ATMP.2009.v13.n3.a5}{\emph{Adv. Theor. Math.
  Phys.} {\bf 13} (2009) 721--896},
  [\href{https://arxiv.org/abs/0807.3720}{{\tt 0807.3720}}].

\bibitem{DHoker:2007zhm}
E.~D'Hoker, J.~Estes and M.~Gutperle, \emph{{Exact half-BPS Type IIB interface
  solutions. I. Local solution and supersymmetric Janus}},
  \href{http://dx.doi.org/10.1088/1126-6708/2007/06/021}{\emph{JHEP} {\bf 06}
  (2007) 021}, [\href{https://arxiv.org/abs/0705.0022}{{\tt 0705.0022}}].

\bibitem{DHoker:2007hhe}
E.~D'Hoker, J.~Estes and M.~Gutperle, \emph{{Exact half-BPS Type IIB interface
  solutions. II. Flux solutions and multi-Janus}},
  \href{http://dx.doi.org/10.1088/1126-6708/2007/06/022}{\emph{JHEP} {\bf 06}
  (2007) 022}, [\href{https://arxiv.org/abs/0705.0024}{{\tt 0705.0024}}].

\bibitem{Assel:2012cj}
B.~Assel, C.~Bachas, J.~Estes and J.~Gomis, \emph{{IIB Duals of D=3 N=4
  Circular Quivers}},
  \href{http://dx.doi.org/10.1007/JHEP12(2012)044}{\emph{JHEP} {\bf 12} (2012)
  044}, [\href{https://arxiv.org/abs/1210.2590}{{\tt 1210.2590}}].

\bibitem{ArkaniHamed:2001ie}
N.~Arkani-Hamed, A.~G. Cohen, D.~B. Kaplan, A.~Karch and L.~Motl,
  \emph{{Deconstructing (2,0) and little string theories}},
  \href{http://dx.doi.org/10.1088/1126-6708/2003/01/083}{\emph{JHEP} {\bf 01}
  (2003) 083}, [\href{https://arxiv.org/abs/hep-th/0110146}{{\tt
  hep-th/0110146}}].

\bibitem{Polchinski:2009ch}
J.~Polchinski and E.~Silverstein, \emph{{Dual Purpose Landscaping Tools: Small
  Extra Dimensions in AdS/CFT}},  \href{https://arxiv.org/abs/0908.0756}{{\tt
  0908.0756}}.

\bibitem{Tsimpis:2012tu}
D.~Tsimpis, \emph{{Supersymmetric AdS vacua and separation of scales}},
  \href{http://dx.doi.org/10.1007/JHEP08(2012)142}{\emph{JHEP} {\bf 08} (2012)
  142}, [\href{https://arxiv.org/abs/1206.5900}{{\tt 1206.5900}}].

\bibitem{Gautason:2015tig}
F.~F. Gautason, M.~Schillo, T.~Van~Riet and M.~Williams, \emph{{Remarks on
  scale separation in flux vacua}},
  \href{http://dx.doi.org/10.1007/JHEP03(2016)061}{\emph{JHEP} {\bf 03} (2016)
  061}, [\href{https://arxiv.org/abs/1512.00457}{{\tt 1512.00457}}].

\bibitem{Alday:2019qrf}
L.~F. Alday and E.~Perlmutter, \emph{{Growing Extra Dimensions in AdS/CFT}},
  \href{http://dx.doi.org/10.1007/JHEP08(2019)084}{\emph{JHEP} {\bf 08} (2019)
  084}, [\href{https://arxiv.org/abs/1906.01477}{{\tt 1906.01477}}].
  

  
  

\end{thebibliography}

\end{document}